\documentclass[a4paper,fleqn]{cas-sc}

\usepackage[numbers,sort&compress]{natbib}

\usepackage{amsthm}
\newtheorem{axiom}{Axiom}[subsection]

\usepackage{subfig}

\usepackage{soul}
\soulregister\cite7
\soulregister\ref7
\soulregister\eqref7
\sethlcolor{white}

\ExplSyntaxOn
\keys_set:nn { stm / mktitle } { nologo }
\ExplSyntaxOff

\begin{document}
\let\WriteBookmarks\relax
\def\floatpagepagefraction{1}
\def\textpagefraction{.001}
\shorttitle{Deep learning in ultrasonic NDE inspection}
\shortauthors{S. Cantero-Chinchilla et~al.}

\title [mode = title]{Deep learning in automated ultrasonic NDE -- developments, axioms and opportunities}

\author{Sergio Cantero-Chinchilla}[orcid=0000-0002-6235-5304]
\cormark[1]
\ead{sergio.canterochinchilla@bristol.ac.uk}
\credit{Conceptualization, Methodology, Formal analysis, Visualization, Writing - Original Draft}

\address{Department of Mechanical Engineering, University of Bristol, Bristol, BS8 1TR, UK}

\author{Paul D. Wilcox}[orcid=0000-0002-8569-8975,]
\ead{p.wilcox@bristol.ac.uk}
\credit{Conceptualization, Writing - Review \& Editing, Resources, Project administration}

\author{Anthony J. Croxford}[orcid=0000-0003-1377-2694,]
\ead{a.j.croxford@bristol.ac.uk}
\credit{Conceptualization, Writing - Review \& Editing, Supervision}

\cortext[cor1]{Corresponding author}

\begin{abstract}
The analysis of ultrasonic NDE data has traditionally been addressed by a trained operator manually interpreting data with the support of rudimentary automation tools. Recently, many demonstrations of deep learning (DL) techniques that address individual NDE tasks (data pre-processing, defect detection, defect characterisation, and property measurement) have started to emerge in the research community. These methods have the potential to offer high flexibility, efficiency, and accuracy subject to the availability of sufficient training data. Moreover, they enable the automation of complex processes that span one or more NDE steps (e.g. detection, characterisation, and sizing). 
There is, however, a lack of consensus on the direction and requirements that these new methods should follow. These elements are critical to help achieve automation of ultrasonic NDE driven by artificial intelligence such that the research community, industry, and regulatory bodies embrace it.
This paper reviews the state-of-the-art of autonomous ultrasonic NDE enabled by DL methodologies. 
The review is organised by the NDE tasks that are addressed by means of DL approaches.
Key remaining challenges for each task are noted. 
Basic axiomatic principles for DL methods in NDE are identified based on the literature review, relevant international regulations, and current industrial needs. By placing DL methods in the context of general NDE automation levels, this paper aims to provide a roadmap for future research and development in the area.
\end{abstract}


\maketitle

\section{Introduction}\label{sec:int}

The automation of ultrasonic NDE processes has been typically restricted to rudimentary tools that give support to a trained operator manually interpreting NDE data. 
An exception can be found in some mass production systems where the NDE for simple inspections of parts with precisely known geometry is fully automated.
Examples of the rudimentary tools include a thickness gauge, whereby the wall thickness is automatically calculated from ultrasonic \mbox{\hl{A-scan}} data~\cite{deutsch2008automated}, or an automated flaw detector that identifies the presence of new echoes over a prescribed amplitude~\cite{markov2003ultrasonic,willcox2003brief}. 
These methods only work on signals with well-defined characteristics; hence they are unable to deal with complex data variations resulting from environmental changes or manufacturing uncertainty and geometric and material complexities.
It is in this context that machine learning (ML) methods show a great advantage as they are based on data and are able to model complex behaviour with high efficiency. 
{Within ML, deep learning (DL) techniques stand out for their capability to extract higher-level features from data using multiple internal layers of processing.}
The move to increasingly digital working practices, in what is widely referred to as Industry 4.0, will produce major growth in the numbers of large datasets over the next few years.
These data will be key for the development of new ML-based inspection processes that match the new opportunities that Industry 4.0 provides, including big data and autonomous systems~\cite{liu2020nondestructive, sophian2021non}.
Note that ML technologies are nowadays relatively inexpensive to develop and easy to implement. Moreover, they potentially enable higher levels of automation of repetitive NDE processes due to their enhanced ability to handle complex scenarios on their own. Nonetheless, this technology is still in early stages of development and application, especially for its application to NDE in safety-critical industrial environments.

This paper reviews the current state-of-the-art of autonomous ultrasonic NDE principally enabled by DL methods. 
The end goal of DL in NDE is twofold: to increase safety levels and to reduce inspection time and cost. These goals are arrived at through (i)~the full automation of processes involving data, (ii)~reduced human intervention in the decision-making chain, and (iii)~the ultimate replacement of human inspectors.
This paper provides a critical analysis of the most relevant and recent scientific contributions in the area of DL-based NDE. The reviewed papers are organised around the NDE tasks addressed -- i.e. data pre-processing, detection, characterisation, and property measurement.
From the review and by comparison with other industrial sectors, a series of automation levels are proposed as a potential automation roadmap, which are analogous to the automation levels used in other domains, such as autonomous aircraft~\cite{EASA2020Roadmap,EASA2021concept}.
To clearly differentiate the previous steps in the automation path, these levels also suggest the main responsibilities to be held by the NDE operator and the autonomous system as systems advance. It is expected that this differentiation will help industries and researchers to categorise their autonomous systems, providing a clearer path to higher automation levels.
Note that the focus of this paper is on ultrasonic NDE and although the levels are generic for any NDE modality, DL contributions to non-ultrasonic NDE are out of the scope of this review. Therefore, different NDE modalities and technologies might currently be at different automation levels. 

Despite the large number of recent contributions in the area, there is still a lack of consensus about the basic properties that DL methods should have for NDE. These properties are different from traditional inspection methods as DL approaches are mostly driven by data rather than physics-based models. 
In the current paper, a set of axiomatic properties for NDE DL are proposed. Like the automation levels, these axioms have been arrived at after discussion with important industries from multiple sectors including aerospace, nuclear energy, renewable energies, and oil \& gas from both manufacturing and operational inspection perspectives. The current implementation level of DL-based automation technology in NDE is relatively low among most of the industries; however, all of them are planning the future implementation of this technology. Moreover, industries are carrying out incipient tests to assess the benefits of DL to enable higher automation for inspection and manufacturing processes.

The outline of this paper is as follows: Section~\ref{sec:dl} introduces some of the most well-known DL architectures. Section~\ref{sec:rev} reviews the most important contributions from the literature. 
Section~\ref{sec:lev} proposes an NDE automation path whereby different automation levels as well as their scopes and implications are discussed. It also identifies remaining challenges and proposes future work within each of the levels. 
Section~\ref{sec:axi} proposes general axioms applicable for the development of NDE DL approaches for any level of automation. It also maps the proposed axioms onto some of the reviewed papers to illustrate the current application of these properties and highlight gaps. Finally, Section~\ref{sec:conc} summarises the paper and selects the most important findings and future directions identified from the literature review.

\section{Neural networks and models}\label{sec:dl}
\begin{figure}[htb]
    \centering
    \subfloat[]{
    \includegraphics{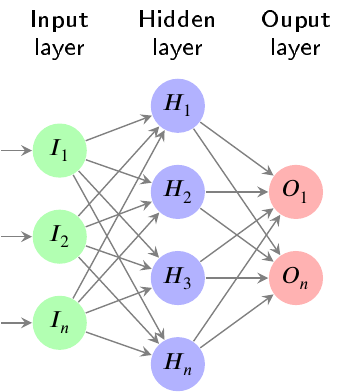}\label{fig:DLarch_a}}\hspace{.3cm}
    \subfloat[]{
    \includegraphics{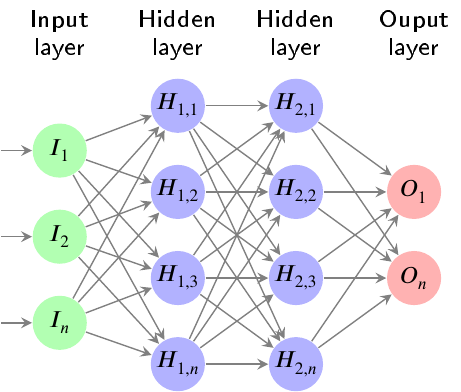}\label{fig:DLarch_b}}\hspace{.3cm}
    \subfloat[]{\scalebox{.9}{
    \includegraphics{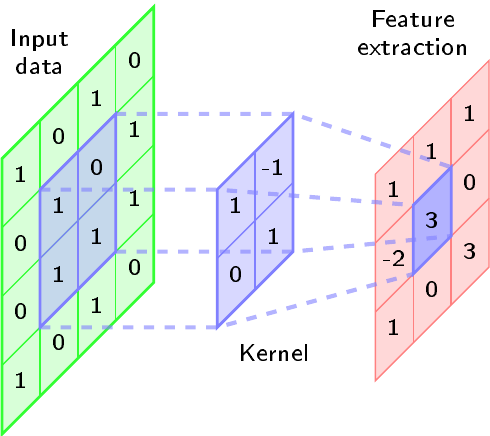}}\label{fig:DLarch_c}}\\
    \subfloat[]{
    \includegraphics{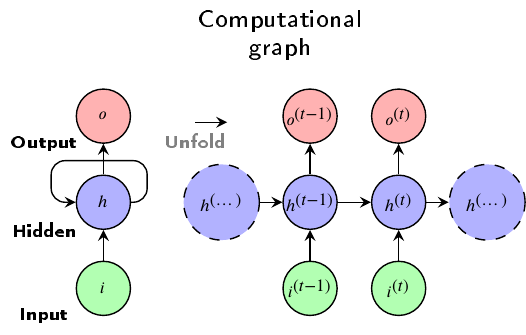}\label{fig:DLarch_d}}\hspace{.1cm}
    \subfloat[]{
    \includegraphics{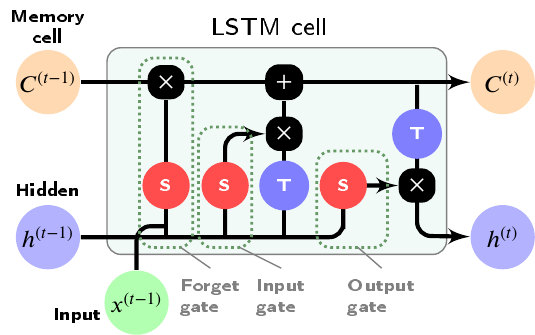}\label{fig:DLarch_e}}\hspace{.1cm}
    \subfloat[]{
    \includegraphics{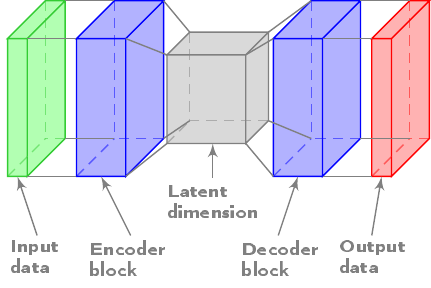}\label{fig:DLarch_f}}
    \caption{Graphical representation of the following DL architectures: (a)~shallow FCNN, (b)~deep FCNN, (c)~CNN, (d)~RNN, (e)~LSTM cell, and (f) classical AE. In these diagrams, green represents input data, pink represents output data, and other colours represent internal data.}
    \label{fig:DLarch}
\end{figure}

Some of the most-used neural networks and already-established DL models are briefly described and presented in this section. Note that the term \emph{network} refers to the general architecture of the DL algorithm, while \emph{model} refers to a specific architecture implementation that is fixed for a purpose. The aim of this section is to give a global view about the functioning of these networks to provide context for the literature review. 
The networks and models, graphically depicted in Figure~\ref{fig:DLarch} and summarised in Table~\ref{tab:DLmodels} (where general overview references are also listed), are as follows:

\begin{itemize}
    \item \textbf{Fully-connected neural network (FCNN)}. This type of feed-forward network is formed by multiple layers that are composed of a number of artificial neurons. These neurons are interconnected with all the neurons in the next and previous layers. The neurons perform the data transformation operations whereby the data $x$ is transformed by a matrix weight $W$ multiplication and a bias $b$ through an activation function $\sigma\left(\cdot\right)$, i.e. $\sigma(W\cdot x + b)$. Two types of FCNNs can be differentiated depending on the number of layers:
    \begin{itemize}
        \item Shallow FCNNs that are comprised of three fully-connected layers (see Figure~\ref{fig:DLarch_a}), namely an input layer with inputs $\{I_1,I_2,\dots\}$, a hidden layer with neurons $\{H_1,H_2,\dots\}$, and an output layer with cells $\{O_1, O_2, \dots \}$. Shallow FCNNs have been used for multiple purposes such as classification and regression~\cite{jain1996artificial}. However, the ``shallowness'' of its architecture makes them relatively limited when dealing with highly nonlinear data and extracting spatial patterns from structured data.
        \item Deep FCNNs, also known as multilayer preceptrons (MLPs), are comprised of multiple fully-connected layers (see Figure~\ref{fig:DLarch_b}). Compared to shallow FCNNs, they add more learning capability for the network to extract non-linear patterns from the data; nonetheless, deep FCNNs are inefficient when dealing with structured data (e.g. 1D time-series, 2D or 3D images) as no account is taken of the relative position of the inputs relative to one another.
    \end{itemize}
    \item \textbf{Convolutional neural network (CNN)}. This kind of network is especially well-suited for dealing with structured data; convolutional layers are able to extract multidimensional features by convolving generally 1D, 2D, and 3D input data into feature maps. Initially designed to deal with visual information, CNNs can be applied to any sort of structured data, such as 1D, 2D, 3D, and even 4D structures of data. They are comprised of convolutional layers that perform a convolution between a kernel and the input data, computing the dot product between the two as schematically shown in Figure~\ref{fig:DLarch_c}. Note that the same kernel weights are used across the input data for each filter. The only requisite for a network to be labelled as a CNN is the inclusion of at least one convolutional layer~\cite{goodfellow2016deep}. CNNs have been applied to semantic segmentation (i.e. classification of image pixels using different labels such as cats, grass or sky), object detection and image classification~\cite{alzubaidi2021review}, among multiple applications. Due to their flexibility and good performance, generic families of CNN models are continuously being developed. Some of the most popular models are:
    \begin{itemize}
        \item AlexNet~\cite{krizhevsky2012imagenet} is a well-known deep CNN model that has achieved success in image recognition and classification.
        \item Visual Geometry Group (VGG)~\cite{mei2021visual} has been widely used for deep image classification and localisation problems.
        \item ResNet~\cite{he2016deep} (standing for residual network) combines the input from the previous layer with the output of the current one. This shortcut reduces the vanishing gradient problem that very deep networks face (the gradient is so small that the weight coefficients cannot change), which enables relatively low prediction errors. ResNet generally has more hidden layers than AlexNet or VGG and has been widely applied to computer vision tasks. 
        \item Inception~\cite{szegedy2016rethinking} contains "inception" modules which perform convolution to input data using multiple different kernel sizes to account for image features with large size variation. Note that there exists multiple version of Inception with enhanced performance achieving better results with less resources, including InceptionTime~\cite{fawaz2020inceptiontime} for 1D processing; Inception-v4~\cite{szegedy2017inception} is the state-of-the-art.
        \item DenseNet~\cite{huang2017densely} was designed to tackle the vanishing gradient problem and builds on the concept of ResNet. It introduced the connection between multiple layers to improve the information transmission.
        \item EfficientDet~\cite{tan2020efficientdet} is the state-of-the-art object detector model that has proven consistently good results in computer vision tasks.
    \end{itemize}
    \item \textbf{Recurrent neural network (RNN)}. This type of network is designed to extract dynamic information about temporal sequential data. RNNs may have different modes depending on how many inputs and outputs make such as one-to-one, one-to-many, many-to-one, and many-to-many~\cite{manaswi2018rnn}. An example of a many-to-many RNN is shown in Figure~\ref{fig:DLarch_d}, whereby the network makes predictions $o^{(t)}$ at the time $t$ based on input data $i^{(t)}$ and the hidden states $h^{(t-1)}$ and $h^{(t)}$; this type of RNN is therefore recurrent on the hidden layer. Note that RNNs can have any number of stacked hidden layers. The classical variants of these networks present some limitations in terms of the information they can store in their memory, which is short-term. This prevents them from making predictions of a quantity which has different frequencies of variation throughout the time, i.e. long-term dependencies.
    \item \textbf{\hl{Long short-term memory (LSTM) and gated recurrent unit (GRU) networks}}. These networks are variants of RNNs that effectively address the main issue (short-term memory) of the classical RNNs, and add functionality to model longer dependencies through a memory cell ($C^{(t-1)}$) and a series of additional gates, such as forget, input, and output gates; see Figure~\ref{fig:DLarch_e} for a graphical description of LSTM cells.  This enables better prediction when the temporal data evolves in a complex manner with multiple frequencies of variation taking place. Similarly, networks based on \hl{GRU}, which are another variant of RNNs, use a single unit to control both the forget gate and the decision to update the state unit. LSTM and GRU networks have been applied to handwriting recognition, speech recoginition and image captioning.
    \item \textbf{Autoencoder (AE)}. This is a particular term is used to describe a network comprised of two separate actors (see Figure~\ref{fig:DLarch_f}): an encoder network that typically extracts representative information (latent representation) from the input data and decreases its dimensionality; and a decoder network that often needs to reconstruct the input (or a variation of the input) data from the latent variables. In such a manner, the network can be trained without supervision to extract representative features from the data. This architecture has also been used for denoising images and time series whereby the input is the noisy data and the output is the denoised data. Note that there is no constraint on the type of encoder and decoder network, so it can be formed by fully-connected, convolutional, and/or recurrent layers for that matter. Probably the most well-known AE model in the literature is the U-Net~\cite{ronneberger2015u} which has been widely used for image segmentation and object classification.
\end{itemize}

It is worth highlighting that any of these base network architectures can be mixed and matched to create other architectures that may provide capacity above and beyond any one of the aforementioned. An example is the Convolutional LSTM network wherein the LSTM cell is converted to perform convolutions rather than modify its state using fully-connected subnetworks~\cite{shi2015convolutional}.

\begin{table}[width=1\linewidth,cols=4,pos=h,h]
\caption{Summary of some of the most used DL architectures and models.}\label{tab:DLmodels}
\begin{tabular*}{\tblwidth\footnotesize}{@{} p{1cm}p{9cm}p{3.5cm}L@{} }
\toprule
Name & Description & Applications & Refs\\
\midrule
Shallow FCNN & Network comprised of one input, one hidden, and one output layers and an undetermined number of artificial neurons & Classification, Data characterisation & \cite{jain1996artificial}\\
\rule{0pt}{3ex}Deep FCNN & Deep network comprised of multiple hidden layers of artificial neurons & Classification, regression & \cite{goodfellow2016deep}\\
\rule{0pt}{3ex}CNN & Network specialised in extracting multidimensional information from structured data & Computer vision, object detection, classification & \cite{goodfellow2016deep, alzubaidi2021review}\\
\rule{0pt}{3ex}RNN & Recurrent network for sequential data whose prediction is based on the previous time step & Prediction, machine translation, image labelling & \cite{goodfellow2016deep}\\
\rule{0pt}{3ex}LSTM & Recurrent network with long-term memory to account for long dependencies in the data & Prediction, machine translation, speech recognition & \cite{goodfellow2016deep}\\
\rule{0pt}{3ex}AE & Network architecture comprised of an encoder and a decoder that extracts representative information from the data & Denoising, feature extraction, image segmentation & \cite{goodfellow2016deep}\\
\bottomrule
\end{tabular*}
\end{table}

\section{Review of the literature}\label{sec:rev}

The main contributions and state-of-the-art of DL based NDE are thoroughly analysed in this section. These are organised considering the NDE task being addressed from data pre-processing (including denoising and imaging) to defect characterisation. 
The most common challenges faced when developing NDE-focused DL methodologies are identified and the solutions illustrated.
Note that the categorisation of this section inevitably causes overlapping within ML tasks for a given NDE problem -- e.g. defect detection can be addressed by binary classification or object detection. Nonetheless, an overall mapping between NDE tasks and the ML tasks found in the literature review is provided in Table~\ref{tab:tasks}.

\begin{table}[width=1\linewidth,cols=4,pos=h,h]
\caption{Mapping between NDE tasks and ML tasks.}\label{tab:tasks}
\begin{tabular*}{\tblwidth\footnotesize}{@{} p{5cm}p{9cm}L@{} }
\toprule
NDE task & ML tasks\\
\midrule
Data pre-processing & Segmentation, detection, encoding, regression\\
\rule{0pt}{3ex}Defect detection & Detection, classification, segmentation\\
\rule{0pt}{3ex}Defect characterisation & Classification, regression\\
\rule{0pt}{3ex}Property measurement & Regression\\
\bottomrule
\end{tabular*}
\end{table}

\subsection{Data pre-processing}\label{sec:rev_proc}
Data pre-processing encompasses techniques that aim to improve the quality of data used for later processes that extract NDE information, but does not usually provide NDE information itself.
In the case of ultrasound data, pre-processing includes but is not limited to: denoising of A-scans (individual signals) and images; feature identification and extraction that allow data reduction; data compression to reduce the communication overhead in applications such as array imaging; and image creation and processing.

\subsubsection{Denoising of A-scans/images}
\begin{figure}[b!]
    \centering
    \subfloat[]{
    \includegraphics{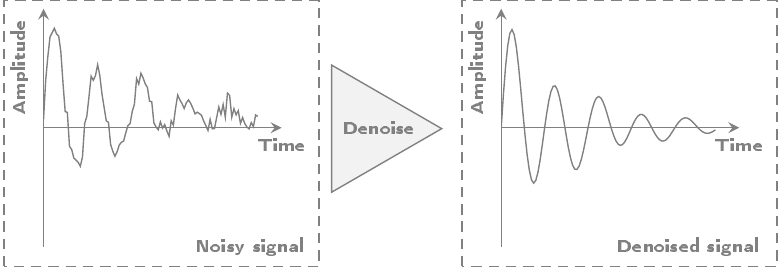}\label{fig:denoising_a}}\hspace{.1cm}
    \subfloat[]{
    \includegraphics{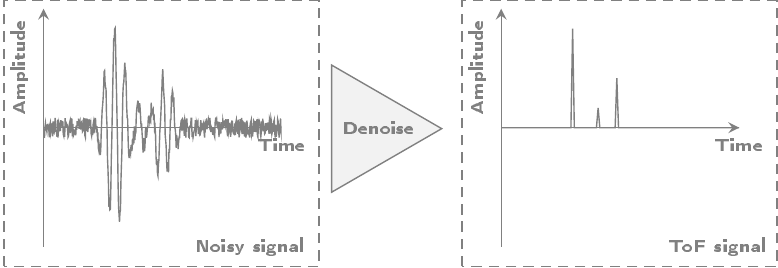}\label{fig:denoising_b}}\\
    \subfloat[]{
    \includegraphics{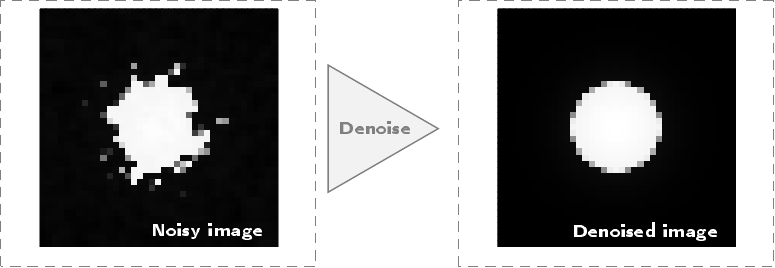}\label{fig:denoising_c}}
    \caption{Examples of different denoising types according to their objective: (a)~denoising raw signals~[22,23,27], (b)~denoising while extracting signal features~[24,25,27], and (c)~denoising ultrasonic images~[29].}
    \label{fig:denoising}
\end{figure}

This section gathers different contributions aimed at suppressing noise, improving signal-to-noise ratio (SNR), and removing selected elements of ultrasonic signals such as boundary-originated echoes. 
In the context of A-scan data and denoising of raw data, Munir~\emph{et al.}~\cite{munir2020performance} proposed a denoising AE in order to improve SNR while enhancing defect classification as a by-product. This DL architecture aims to extract basic noise-free features from ultrasonic signals and reconstruct almost noise-free data (see Figure~\ref{fig:denoising_a}). Two separate datasets were used for training and testing purposes with 3825 and 2100 signals in each dataset. A data augmentation strategy through time-shifting of experimental signals was applied to create more training data from smaller sets of experimental signals. In comparison with classical methods for suppressing high levels of noise such as bandpass filters and averaging, the proposed denoising AE provides a computationally efficient approach that is highly generalisable. The classification performance was improved between 1-10\% depending on the defect type.
Gao \emph{et al.}~\cite{gao2020ultrasonic} proposed a hybrid method for denoising ultrasonic signals using image-based AEs. The goal was to obtain a signal processing method for denoising without the need for input parameters, hence being less prone to human errors. The main idea was to map the signal from an array of values into a time-amplitude image of the signal, which is then fed into an AE that denoises the signal by correcting the location of pixels. The denoised image is then transferred back into a time domain signal with higher SNR. The training and testing datasets consisted of 500 and 100 experimental signals from an ultrasonic phased-array configuration, respectively. In comparison with other denoising methods such as the empirical mode decomposition (EMD), principal component analysis (PCA) and singular value decomposition~(SVD), the proposed approach adds independence on the definition of signal characteristics and makes no assumption about decomposition layers or wavelet basis functions. The denoising results demonstrated robustness across different SNR levels.

Another type of denoising is the clean extraction of a value of interest (e.g. time of flight (ToF)). An example of this denoising type can be found in~\cite{chapon2021deconvolution}, where the authors proposed a method for improving the balance to be struck between penetration depth and axial resolution, especially when the frequency increases leading to higher attenuation and lower SNR. A CNN was designed for separating overlapping echoes from reflectors and boundaries, while estimating the ToF and amplitude of each echo as schematically shown in Figure~\ref{fig:denoising_b}. This has been typically addressed using cross-correlation or deconvolution. The proposed approach overcomes some limitations such as the need for prior knowledge of interface reflectivity and multiple reflection paths and the sensitivity to noise. The DL method was trained using 2000 simulated signals from a finite element model with defect and its performance was demonstrated using experimental data. The results reported a high success rate of detected reflectors and very low false detection rate up to a SNR of 20dB, above which the accuracy degrades quite drastically.
Cantero-Chinchilla \emph{et al.}~\cite{cantero2021deep} proposed DL architecture for identifying and suppressing structural artefacts from full matrix capture (FMC) data, leading to clearer ultrasonic images obtained through the multi-view total focusing method (TFM). An AE-based approach was developed consisting of an encoder that extracts physical parameters (e.g. specimen thickness, probe angle, and probe stand-off) from the FMC data and a decoder that takes as input the encoder output and predicts the arrival times of the structural artefacts. The arrival time information is used to suppress the artefacts by applying masking windows in the original FMC data. Then, the masked data are used for ultrasonic multi-view imaging and compared with an approach to remove artefacts directly in the TFM views~\cite{bevan2018experimental}. A simulated dataset of 4913 defect-free FMCs were used to train the DL models while the validation was performed on experimental data. The results demonstrate superior performance of the DL approach for removing artefacts without completely masking out areas of the image, hence letting through more information about possible defects.
Gao \emph{et al.}~\cite{gao2021domain} developed an autonomous system for enhancing ultrasonic logging through (1)~denoising raw pulse-echo data, and (2)~ToF estimation. Compared to existing methods which include model-based methods (e.g. maximum likelihood estimation and expectation maximisation), prior knowledge deconvolution, and local-global window, the proposed DL framework proves to be more efficient, less knowledge-dependent, and more accurate in noisy environments. The proposed denoising method consists of two DL networks, a decoder-U-Net architecture and an AE for denoising and reconstruction purposes. The encoded features are used as input for a FCNN for the ToF. The authors used \textit{transfer learning}~\cite{tan2018survey} from microseismic P-wave (labelled) signals to unlabelled pulse-echo data, which in turn improves the ToF detection accuracy due to the extraction of shared hidden features from both datasets. Transfer learning addresses the issue of insufficient training data by training a DL model on a source domain and transferring the knowledge to a different target domain. The source domain consisted of expert-labelled field microseismic data while the target domain was comprised of unlabelled samples obtained in laboratory and on the field.

An example of image (C-scan) denoising can be found in~\cite{jedrusiak2020deep}, where the authors developed a DL approach for denoising air-coupled ultrasound data, which are well-known for having low SNR, that improves the performance of classical filters (e.g.~Gaussian function-fitting or the Savitzky-Golay filter). In this context, C-san data were denoised in two steps: (1)~a LSTM model combined with 1D convolutional layers to classify each pixel (or signal in the scan) in two states, i.e. damaged or undamaged; and (2)~the binary image from (1) is then fed into a CNN model that removes most of the image noise leading to a cleaner and more accurate representation of the defect (see Figure~\ref{fig:denoising_c}). The training dataset for (1) consisted of 369,730 experimental data points (i.e. from individual signals in laboratory C-scans) and 109,370 for (2). The main expected benefit of this approach was to make air-coupled ultrasound more useful, improving the quality of the data. The method provided a high level of classification accuracy (over 90\%) with misclassification mainly located at the defect edges.

From the reviewed papers, it is worth highlighting AEs for denoising applications. One of the main benefits of using AEs is that they do not require any prior assumptions and enable more accurate defect classification. However, AEs can lose a certain degree of ultrasonic information, especially from low-amplitude echoes. This information loss may be critical for some applications such as defect characterisation. Alternative training strategies (e.g. through generative adversarial networks~\cite{creswell2018generative}) and increased model complexities could be adopted to address this issue. Another issue is the lack of real training data, which has been addressed in many cases by simulation and in~\cite{gao2021domain} through transfer learning. This is especially relevant to NDE applications, where large experimental databases are still lacking, particularly those with damage.

\subsubsection{Data reduction and compression}
A critical pre-processing step for many defect detection and/or classification techniques is the extraction of meaningful features. AE models have been used for data reduction purposes in applications other than ultrasonic NDE such as guided-wave and image-based structural health monitoring. The key idea is that, given the inherent structure of AEs, the input data can be compressed into a few latent features that represent the original data that are learned in an unsupervised manner by the DL model. These features can then be used for anomaly detection~\cite{mao2021toward} by comparing the distance between the extracted features of the input measurements. In this work, the anomaly detection method was trained using 10,000 data samples (images from experimental acceleration time-series of a long-span bridge) and 7000 for testing purposes. In comparison with classical feature extraction based on statistical measures such as cross-correlation, mean, standard deviation, and other statistics, DL-based autonomous feature extraction adds more flexibility, leaving the algorithm to decide how the feature is best described and what value should be assigned. The main shortcomings to this approach are (1)~the relatively large amount of training data required and (2)~the lack of physical meaning of the features which are difficult, if not impossible, to interpret. 

In the ultrasonic NDE realm, there are a relatively small number of publications dealing explicitly or implicitly with feature extraction. An example is the paper by Hong~\emph{et al.}~\cite{hong2019liquid}, where the main aim is to detect liquid level in porcelain bushing type terminals (cylindrical pipe-like elements used for high voltage insulation). The guided-waves are obtained by attaching transducers to the wall of the insulating element and propagating through multi-path transmission in the multi-layered media comprised of air, ceramics, and oil. Here, guided-wave signals are divided into multiple segments which are wavelet-transformed into time-frequency representations. These data are then introduced into an AE for feature extraction (through the encoder). It is important to note that the main expected benefit of this approach is the reduction of expert dependency in interpreting complex data. The features are then used as input into a regressor neural network for inferring liquid level. The AE and regressor networks were trained using experimental guided-wave data from a porcelain bushing type terminal with different liquid levels. This type of DL-based feature extraction method is compared against principal component analysis (PCA), with the proposed AE model showing superiority in terms of accuracy. 
Note, however, that a fully-linear AE approximates PCA, and that the relative superiority of general AEs comes from the non-linearities typically used in deep networks -- i.e. non-linear activation functions such as sigmoid and ReLU.
Therefore, once again the flexibility and suitability of AE models when dealing with ultrasonic data especially in the pre-processing stage is demonstrated.
    
Data transmission may become a bottleneck in some data-intensive applications such as ultrasonic imaging through array probes with high numbers of elements.~To alleviate this issue, a few contributions can be found on data compression. For instance, Pilikos~\emph{et al.}~\cite{pilikos2020deep} proposed an AE structure for data compression of ultrasonic array signals to later reconstruct them and create ultrasonic images. The method aims to enhance data transmission rates by enabling higher compression rates than classical approaches such as compressive sensing. The model comprises a 3D CNN encoder whose output is mapped into a series of discrete latent variables via nearest neighbour search. The training and testing databases consist of 700 and 70 sets of simulated data. The latent information is then decompressed using a 3D CNN decoder model and imaged using a delay and sum (DAS) algorithm. All the elements (encoder, decoder and imaging) are implemented together in the training strategy, which achieves a better performance than training the AE on its own.
In 2015, Kesharaju~\emph{et al.}~\cite{kesharaju2015feature} proposed a feature selection methodology based on genetic algorithms (GA) and FCNNs whereby a few pre-selected features are used as input of the FCNN model for classifying ultrasound signals from defects in complex ceramic materials. The training datasets for the DL method consisted of 132 experimental signals. A comparison with PCA was performed, showing faster and slightly improved performance in the GA and FCNN methodology.
Both of these examples highlight the benefits of using DL for data compression, with the AE model structure being probably more appropriate due to its inherent bottleneck architecture.

\subsubsection{More efficient / higher resolution image formation}
This area is one of the most prominent in ultrasonic NDE, given the inherent need for images to ease the interpretation of complex ultrasonic signals and because this is well-suited to existing DL image processing models (see Section~\ref{sec:dl}). An example of the application of DL methods for imaging is the one provided by Pilikos~\emph{et al.}~\cite{pilikos2020fast} where an end-to-end DL model for data pre-processing, image formation, and image post-processing was proposed. This method aims at reducing errors (e.g. from inaccurate physics modelling or noise in the data) that propagate and are possibly amplified during the imaging process. The raw data are filtered through a 3D CNN and then introduced into the DAS algorithm for image formation and lastly a 2D CNN is used for image segmentation. The networks are trained together by allowing the backpropagation algorithm to link and correct errors in both networks for the image segmentation objective. A total of 180 and 50 samples are simulated for training and validation purposes, respectively. Note also that in their work, the authors used the DAS algorithm as an intermediate layer of the DL network. The performance of the networks jointly trained was demonstrated to be higher than those trained independently.
Alguri~\emph{et al.}~\cite{alguri2019transfer} proposed an AE for the reconstruction of wavefield images from a reduced sample dataset from laser Doppler vibrometer (LDV). This work is motivated by the relatively large amount of time required to acquire full wavefield data from LDV scans, hence using a reduced subset of measurements can lead to a more efficient procedure. The proposed method applies a transfer learning strategy to adapt the model initially trained on comprehensive simulated data (2500 data samples) to then train the encoder part on sparse data, fixing the weights of the decoder. This strategy allows consistent behaviour of the encoder with both complete and sparse datasets, while ensuring optimal functioning of the imaging (the decoder) part of the network. As a result, the DL model was able to recover most of the wavefield images.
Keshmiri~\emph{et al.}~\cite{keshmiri2020deep} developed a DL approach for reducing measurement points for wavefield imaging for high-resolution images. The expected benefits of this approach were to achieve higher compression rates while providing greater resolution than other methods such as compressive sensing~\cite{di2015compressive}. The data are firstly compressed to reconstruct the image with an evident degradation of its resolution. The reconstructed image is then passed through the CNN model to achieve a higher resolution similar to the uncompressed image. A patch-based approach and a total of 326 full wavefield images (i.e. dividing them in smaller images to have 177,687 images) are used for training, validation and testing of the proposed network. The results show better performance of the DL method than classical compressive sensing while being more consistent across different compression rates.
Mei~\emph{et al.}~\cite{mei2021visual} proposed a hybrid VGG-U-Net network for improving ultrasonic image resolution in cases where the surface is curved given the associated limitations such as refraction and multiple internal reflections. DL is used here to overcome the diffraction limit and automate the imaging of complex structures with less expert dependency. The idea of the model is based on a U-Net where the encoder is substituted by the VGG network, which is able to capture more details about the data. The input of the VGG-U-Net is an uncorrected image and the output is the corrected image. The results from simulated data show a clear superiority over track-scan imaging~\cite{mei2020robot} (i.e. imaging method that uses ultrasonic data from a robot scanning a curved surface) with higher resolution. A total number of 1560 simulated ultrasonic images are used for training, validation and testing. Experimental images show good approximates but with less accuracy.
Note that these works are also linked by their motivation to obtain the same results with less data, hence optimising acquisition and processing resources.

On a different imaging application, Song~\emph{et al.}~\cite{song2020super} created a super-resolution DL model for overcoming the diffraction limit and achieving subwavelength resolution using guided waves. The model is expected to provide this super-resolution using phased-arrays and have more robustness to noise than the time-reversal MUSIC (multiple signal classification) algorithm. The proposed model is comprised of a first CNN network that detects defective areas from a TFM image and a second CNN network that locally resolves the fine details of the detected area. The number of simulated training, validation and testing samples is 6144 TFM images for both the detection and super-resolution networks; the number of experimental TFM images is 19439 stemming from 5 aluminium plates and multiple defect types. Note that data augmentation by image manipulation (e.g. rotation or mirroring) was applied to produce a large amount of images. The results evidence a good agreement between the output and ground truth while achieving a greater robustness to noise in comparison with the time-reversal MUSIC algorithm.

It is evident that the use of DL for ultrasonic image processing has particularly benefited from the natural suitability of many DL networks and models for that purpose. The main common architectures are CNNs and AEs that are used for different purposes, such as increasing image resolution or image segmentation. They have proven effective through these papers, but it is worth highlighting the promising results in image super-resolution~\cite{song2020super}, being able to overcome physical limits (e.g. diffraction limit) and enhance the image resolution, thus promoting an easier and more precise defect characterisation and the need for less demanding hardware.

\subsection{Defect detection}\label{sec:rev_det}
This NDE task provides binary information about whether or not a defect is present in the inspected structure. 
Defect detection in A-scans using DL approaches has been addressed by Guo~\emph{et al.}~\cite{guo2019fully}, who proposed an automated classification approach for ultrasonic signals. By using DL, it is possible to detect defects directly from A-scans instead of C-scan images, which gives a performance advantage as A-scans are faster to acquire and generally require cheaper equipment to obtain. The DL model consists of a parallel architecture of GRU and CNNs to process ultrasonic signals (extracting temporal and global features, respectively) and classify them into two classes: defect and no defect. The training, validation, and testing datasets consisted of 3600 ultrasonic signals experimentally acquired. The proposed model provides higher accuracy than other networks such as LSTM, GRU, or ResNet (refer to Section~\ref{sec:dl}).
Yan~\emph{et al.}~\cite{yan2020deep} developed a CNN-SVM (support vector machine) framework for the automated identification of pipeline girth cracking through ultrasonic signals obtained from electromagnetic acoustic transducers (EMATs). The motivation for using DL in this work is to overcome the limitations of using EMATs, i.e.~low transmission efficiency and lift-off leading to low SNR. This ultimately poses difficulties in interpreting data for pipeline girth weld cracks. In this approach, the CNN works as a feature extractor while the SVM is in charge of classifying the data as coming from defective or non-defective structures. The model receives as input the spectrogram of the filtered EMAT signal. A total of 2160 experimental signals were used to train the DL framework. The proposed approach outperforms others whereby the features are extracted with classical methods such as the discrete wavelet transform, the Shannon entropy, and other statistical features. It is then observed how both RNNs and 1D CNNs are used for defect detection through A-scan data as the two DL architectures are well suited to deal with temporal and array-structured data.
    
DL has also been applied to ultrasonic B-scans for defect detection. For instance, Yuan~\emph{et al.}~\cite{yuan2019automatic} proposed a FCNN framework to identify echoes from defects in B-scans from train wheels. ML is proposed here as a way to automate inspections and address data-related heterogeneity stemming from, for example, different background noise levels because of the wheel structure's complexity. The model consists of two networks, the first one classifies defect and noisy signals, while the second one classifies echoes coming from defects or elsewhere. The networks receive manually extracted features from the B-scan data (e.g. variance and peak SNR) as input. A total of 736 and 582 experimental data samples were used to train both networks. The results show a relatively high level of accuracy with the 92\% defect recognition.
Medak~\emph{et al.}~\cite{medak2021automated} proposed an automated defect detection network through a EfficientDet network (originally developed for object detection) that takes as input ultrasonic B-scan images. The proposed approach uses an image processing DL architecture (EfficientDet) that outperforms other DL models (YOLOv3~\cite{redmon2018yolov3} and RetinaNet~\cite{lin2017focal}) having better performance in new data, while still automating the defect detection. B-scan images are fed into the model and the defect echoes are identified through a box overlapped in the defect location, effectively detecting defects. A total of 6637 experimental images containing different defects were used to train the DL framework. The results show better detection accuracy compared to other DL models such as ResNet.
Virkkunen~\emph{et al.}~\cite{virkkunen2021augmented} performed an interesting study whereby the performance of a CNN model for detecting flaws from phased-array ultrasonic B-scan images is compared against the performance of three human operators. The CNN model is trained using data augmentation to overcome the problem of limited data and aid learning. Virtual flaw signals are modified to recreate flaws at different locations, depths, or lengths. A total of 20000 data samples were generated by combining experimental defect-free and simulated defect data to train the CNN. The paper shows better performance of the CNN model than the operators with 0 and between 2-36 false calls, respectively.

Slonski~\emph{et al.}~\cite{slonski2020detection} studied the automation of flaw detection in concrete through ultrasonic tomography images. DL is proposed to enable the automation on the basis that it will be more efficient, hence cheaper, yet less prone to error than operator inspection. The authors used a VGG-16~\cite{simonyan2014very} network, which is pre-trained on a large database of images (ImageNet~\cite{deng2009imagenet}) and fine-tuned by training the last set of convolutional layers using a set of 246 B-scan images taken in the lab (i.e. through transfer learning). The authors reported a validation accuracy of 97\%.

Lastly, it is worth highlighting the work by Ye~\emph{et al.}~\cite{ye2021benchmarking} whereby a comprehensive dataset of ultrasonic wavefield images and a benchmark of some of the most well-known DL models is provided. Although there is no new model proposed in this publication, this work can guide researchers and practitioners to choose the right model that suits their needs. A set of 7000 ultrasonic inspection images from a beamforming-laser emitter-receiver configuration was used in undamaged and 17 different cases of damage forms in stainless steel plates. In particular, the following models are tested: AlexNet, VGGNet, ResNet-18, Inception-v3, Wide Resnet, DenseNet, and SE-ResNet-50~\cite{hu2018squeeze}. Note that these DL models are well-suited for image processing, hence its application for defect detection based on ultrasonic images. The readers are directed to~\cite{ye2021benchmarking} for succinct but clear explanation about these networks. The models are used for detecting defects and they are compared based on the detection accuracy, model complexity, memory usage, and computational efficiency. The most accurate model was DenseNet closely followed by SE-ResNet-50, which is slightly less complex but also computationally efficient.

Furthermore, transfer learning has also proven effective in pre-training DL models in large image databases to then be fine tuned in the smaller ultrasonic databases. In any case, DL methodologies have exceeded the performance of classical methods and even human operators, suggesting that the potential automation of repetitive NDE tasks is plausible in the medium term.

\subsection{Defect characterisation}\label{sec:rev_cha}
Defect characterisation aims to extract NDE information from data and predict the type and numerical characteristics (e.g. crack length) of a defect. The reviewed DL contributions for defect characterisation have been organised depending on their ultimate goal: (1)~to classify defects or (2)~to quantify defect properties (e.g. crack length).

\subsubsection{Defect classification}
Steel plates and welded regions have been the subject of numerous investigations into DL classifiers that are able to differentiate types of defects, e.g. porosity, lack of fusion, lack of penetration, and cracks among other defects. Typically, FCNNs and CNNs are used as network architectures.
An early example of the characterisation of defects is the one provided by Bettayeb~\emph{et al.}~\cite{bettayeb2004improved} in 2004, whereby the authors developed a FCNN for automating the classification of defects from \hl{A-scans}. The proposed network takes as input a signal from defective material and gives as output a binary classification between planar and volumetric defects. A high classification accuracy (over 95\% of recognition rate) was evidenced in their study.
Sambath~\emph{et al.}~\cite{sambath2011automatic} in 2011 proposed a FCNN for automating and increasing sensitivity to the detection and classification of flaws from ultrasonic testing in welds. A wavelet transform was applied to the raw data to extract features that were fed to the FCNN in order to classify the data into four classes: porosity, lack of fusion, tungsten inclusion, and no defect. A total of 240 ultrasonic A-scans were used for training and validating the FCNN. An accuracy of 94\% was obtained proving it to be better than other approaches without signal processing for the raw data.
In~\cite{virupakshappa2019multi}, the authors addressed the detection and classification of flaws using ultrasonic inspections of metallic samples with the ultimate goal of automating this action which was previously performed manually. In this approach, a deeper architecture was proposed by using a B-scan image-driven CNN followed by fully-connected layers that predicted the defect type (e.g. side-drilled hole, flat bottom hole, etc.). The dataset contained 400 simulated B-scan images for all defect types. An accuracy of over 90\% was obtained for all the defect types proving its potential.
In a progression of three works, Munir~\emph{et al.}~\cite{munir2018investigation, munir2019convolutional, munir2020performance} addressed the classification of defects in welds to allow automation while obtaining high levels of precision. The authors started by using a FCNN in 2018~\cite{munir2018investigation} whereby the DL model classified the defective signal into a set of possible defects, such as cracks, porosity or slag inclusion. Later in 2019~\cite{munir2019convolutional}, the authors proposed a CNN followed by fully-connected layers along with data augmentation strategies for increasing the DL model capabilities. Lastly in 2020~\cite{munir2020performance}, an AE was introduced as an initial step to denoise the ultrasonic signals and hence improve the performance of the CNN architecture.

Different materials and structures have also been the subject of investigation for the application of DL methods to ultrasonic NDE data. For instance, Meng~\emph{et al.}~\cite{meng2017ultrasonic} developed a CNN network for the automation of ultrasonic signal classification from C-san signals in a carbon fibre reinforced polymer (CFRP) structure. The network uses as input the wavelet transform of the raw data and accurately gives as output a classification between two types of defects: void or delamination. A total of 6000 experimental A-scan signals were used for training and validation purposes. The paper also shows evidence of this approach working on a CFRP sample with multiple inclusions and locating the different defects with high precision.
Rodrigues~\emph{et al.}~\cite{rodrigues2019carburization} studied the automation and enhancement of carburising level estimation for high pressure steel pipes using ultrasonic inspection. The model consists of a FCNN that takes as input the discrete Fourier coefficients of a raw signal and accurately classifies it into three different categories: high, low, or no carburisation. The number of experimental signals used for training was 200.

DL techniques, mostly comprised of convolutional and fully-connected layers, have been successfully able to classify defects in metallic components, and in particular in welds where the majority of defects are encountered. However, the application of DL to other more complex (anisotropic) materials such as composite plates has been scarcely addressed. These structures pose great challenges from the multiple internal reflections caused by different layers and their anisotropy. Nevertheless, there are DL architectures that are complex enough to extract targeted information from ultrasonic measurements in composite structures. Efficient physics-based models~\cite{bochud2015sparse} for creating relatively large training databases as well as transfer learning will be critical in the success of this application.

\subsubsection{Defect sizing}
Another type of defect characterisation is through the inference of a real number that quantitatively sizes defects, e.g. through crack length estimation. 
Pyle~\emph{et al.}~\cite{pyle2020deep} addressed this issue and also the lack of real (experimental) training data for DL models. The authors used a hybrid FE-ray-based to generate synthetic data (in the form of 25,625 simulated ultrasonic images) to train a DL model of crack characterisation (i.e. length and angle). The DL model is then applied to experimental data with more accurate estimation of both crack angle and length than the classical 6dB drop method. This work demonstrates that the use of physics-based models can boost the performance of DL models.
Miorelli~\emph{et al.}~\cite{miorelli2021defect} proposed a CNN for automating the defect localisation and sizing from ultrasonic guided waves. The input data consist of DAS images from circular piezoelectric transducer layouts and the output of the model is the continuous values of the XY position and radius of the defects in aluminium plates. A total of 500 simulated images were used to train the CNN, while validation and testing were performed on both simulated and experimental images. The DL model was tested with both modelling and experimental data, showing a high correlation with the true values in either case.
Bai~\emph{et al.}~\cite{bai2021ultrasonic} has recently provided a comparative study between classical Bayesian inversion approaches and a CNN regressor model -- i.e. the SMInvNet based on the model developed by Miorelli~\cite{miorelli2021defect}. Here, the input data consist of the scattering matrices (measured from FMC and TFM data) while the output is the size and angle of a surface notch. The size of the training and test sets was 976 and 180 simulated samples, respectively. In this study, the classical Bayesian approach shows a higher degree of characterisation accuracy and less uncertainty, while SMInvNet gives comparable performance but with more dispersion stemming from the model's epistemic uncertainty. It is worth highlighting that from this work, it was concluded that the uncertainty interpretation from the DL model remains an open challenge. In contrast, the Bayesian approach provides a natural and explicit uncertainty map of the unknown defect parameters (at the cost of more expensive inverse problem evaluations).

Interestingly, the use of DL for the regression of defect parameters has not yet been fully exploited. This application can potentially enable a higher automation of the NDE that can systematically evaluate the criticality of small defects. As in previous sections, transfer learning proves to be essential when the issue of lack of defect data are present. Pyle~\emph{et al.}~\cite{pyle2020deep} demonstrated that the use of modelling data can help in training DL models even if they are applied to experimental measurements with high-quality modelling. Notwithstanding this, multiple domain adaptation techniques (e.g. by mixing modelling and experimental databases with different weights in the loss function) need to be explored to ensure their suitability for different ultrasonic NDE applications.

\subsection{Property measurement}\label{sec:rev_prop}

Deep learning has also been used for material characterisation (other than defects), which include the direct or indirect measurement of material properties. For instance, the inference of porosity level (obtained by processing C-scan images) in additive manufacturing parts have been investigated by Park~\emph{et al.}~\cite{park2021porosity,park2021deep}. Several DL models based on CNNs, and FCNNs, were designed and tested against real samples by taking as input the ultrasonic signal and as output the porosity level (through multiple classes). The size of the ultrasonic databases varied between 1000 and 6000 experimental signals.  Out of the different architectures proposed, the convolutional approach proves to be the most effective with less classification error between different porosity intervals.
Ma~\emph{et al.}~\cite{ma2020ultrasonic} proposed a DL-based approach for the estimation of the porosity level in thermal barrier coatings (widely used to protect hot section components of aircraft engines) from ultrasonic A-scans. Ultrasonic reflection coefficients are used as input of the FCNN with three hidden layers and it predicts the porosity characteristics. An optimisation procedure was proposed to achieve the optimum ultrasonic parameters to estimate the porosity and then a Gaussian process is used for the regression of the actual porosity level from each A-scan. Therefore, DL is used here to substitute the classical approach of establishing a calibration curve that relates the ultrasonic parameters with the porosity level. DL provides a faster and more flexible approach to adapt to the nonlinearities of the data. The results show that the optimum set of ultrasonic parameters provides the most accurate porosity estimation.
L{\"a}hivaara~\emph{et al.}~\cite{lahivaara2018deep} studied the feasibility of using CNNs applied to ultrasound tomography for the property estimation of water-saturated porous materials.~Given the large computational burden that solving the forward model of wave propagation in coupled poroviscoelastic-viscoelastic-acoustic media entails, the authors proposed a faster inverse problem through a CNN regressor of material properties. Input data are in the form of B-scan images while the output is comprised of material porosity and tortuosity of the inclusions.~The total number of samples in the training set was 75,000 samples stemming from simulations and data augmentation by adding different levels of noise.~Numerical results show an excellent correlation between predictions and ground truth, while the efficiency of the DL method is higher than a classical inversion approach -- i.e. performing thousands of forward model evaluations.

The estimation of other material properties from ultrasonic data has also been the subject of study. For example, the estimation of the spatial distribution of grain size in metals was investigated in~\cite{dapkus2020study}. The authors proposed a DL architecture consisting of a CNN followed by several fully-connected layers. The input of the model is comprised of B-scan images while the output is the size of the material grain (given as the probability of the data belonging to three different classes based on grain size). The total number of experimental ultrasound signals used in this work was 302,400. The end goal of DL is to enable the estimation of a property difficult to obtain, which typically relies on structural noise distributions.
Singh~\emph{et al.}~\cite{singh2021real} proposed two DL architectures for creating ultrasonic tomographic images with grain orientation information. The first model consists of multiple FCNNs (one per pixel) that take as input the ToF matrices from ultrasonic arrays and provide the grain orientation information as output. The resulting image and additional prior information are then fed into a generative adversarial network to increase the image resolution. A total of 7500 and 6000 simulated datasets were used to train the FCNN and generative adversarial network, respectively. DL is used as a tool that enables fast tomographic imaging with high resolution for an enhanced detection of flaws during manufacturing. The results show that the DL framework efficiently reconstructs the tomographic images from the ToF information and concludes that there is room to improve the high resolution reconstruction accuracy adopting advanced loss functions such as the Wasserstein distance~\cite{arjovsky2017wasserstein}.
Gopalakrishnan~\emph{et al.}~\cite{gopalakrishnan2020deep} compared two different DL approaches for estimating elastic properties of composite laminates from ultrasonic guided wave data. The methods consisted of (1) a CNN followed by fully-connected layers, and (2) a LSTM recurrent network. The total number of data samples used for training was 2719 simulated signals. DL is adopted here to substitute classical deterministic inversion schemes that are computationally slow and prone to noise-induced errors; they are also limited in terms of large-scale automation. Their results suggests that, although both DL approaches are accurate in estimating the elastic properties of composite laminates, the LSTM network performs better with noisy data.

A recent study has addressed the quantification of microstructure properties in polycrystalline Nickel using physics-informed neural networks~\cite{shukla2021physics}. Few existing methods infer microstructure from full wavefield data from the displacement vector, and hence DL is used here as an accurate and efficient inversion method. The authors proposed three networks for the inference of three elements of the stiffness tensor (as polycrystalline Nickel exhibits a cubic symmetry) and uses ultrasonic wavefield data as training data. Hence, two networks are trained when new wavefield data are acquired: (1)~a network to recover the wavefield measurements as a function of the spatial coordinates and the time, and (2)~a network to infer the material properties having to satisfy the underlying governing equation along with the wavefield predicted by (1). Note that the physics part is introduced in the loss function by a combination of mean squared error of the wavefield displacement and the corresponding wave equation (in-plane and out-of-plane) that the predicted stiffness element and displacement must satisfy. The results show a high accuracy predicting the values of the stiffness elements. This study shows the great potential of creating mixed physics- and data-based DL approaches that are able to handle complex problems in a highly efficient manner.
Similarly, in a previous work from the same authors~\cite{shukla2020physics} a physics-informed neural network was proposed to characterise the spatial acoustic velocity distribution. Hence, by combining two networks inferring wavefield displacement and velocity, respectively, this approach proved to be capable of characterising possible defects by visually identifying areas with lesser wave velocity. Again, this work imposed the residual of the governing wave equation as an additional term of the loss function, hence educating the data-driven approach with physics. A relatively small database of 30 snapshots with about 20\% of spatial information was used to train the model.

\subsection{Common challenges in the literature}
From the review of existing work, it is clear that there are multiple shared challenges that are continuously faced when developing DL models for NDE, regardless of end goal. Many of these challenges are ultimately related to the lack of training data. This issue is addressed through different techniques: (1)~data augmentation, (2)~generative models, and (3)~transfer learning (including domain adaptation techniques). Data augmentation consists of creating additional data samples in the training data set by manipulating the original data~\cite{shorten2019survey}. This manipulation may be a simple rotation, scaling, or stretching of original data, which most of the times are images. Data augmentation of 1D signals has also been addressed through the addition of artificial tone-bursts acting as defect reflections for example. Domain adaptation techniques can be used to refine the performance of DL models in different domains to those for which they were originally trained~\cite{wang2018deep, zhao2020multi}. The main example in the NDE context can be found in a scenario whereby a DL model is designed and trained with modelling data and then a small subset of experimental data are used for fine tuning -- hence making the model more accurate in the experimental domain. This technique is particularly useful when a model is available but experimental data are limited as it capitalises on the strengths of well understood physics-based models.
Generative models such as variational autoencoders~\cite{kingma2019introduction} and generative adversarial networks~\cite{creswell2018generative} can also be used to generate data from multiple classes.
Lastly, transfer learning is used to use the latent knowledge of a DL model that has extracted from a source domain to be applicable onto a target domain~\cite{tan2018survey}. For example, an image segmentation DL model that has been trained on photographs of various objects can likely be used as a strong starting point to train a model which can segment ultrasonic images with defect inclusions.
Related data issues are (1)~acquiring and developing the corresponding labelled data for DL model training, especially in tasks like segmentation where extremely detailed labels are required; and (2)~confirming ground truth involving complementary NDE techniques such as X-ray and computed tomography. These issues could be addressed using active learning~\cite{gal2017deep} whereby the algorithm learns from small amounts of data and queries the user to label new data.

Another important challenge that has not yet been addressed in the literature is the lack of a universally-accepted method (e.g. through standard metrics such as probability of detection) for performance quantification. This makes it inherently difficult to compare different DL works even on the same application. For instance, although some authors use direct error estimation from the training procedure to estimate the quality of a DL architecture, others are using distributions that characterise both error and dispersion. 
DL approaches can either be developed to adapt currently-existing standard metrics or be used to create new standards in places where they do not currently exists. Nonetheless, a unified methodology for the quantification of performance would make the DL contributions more meaningful and potentially more attractive to industries as they can be easily compared.

Similarly, the quantification of uncertainties has been barely addressed for the estimation of output variance. A global methodology to estimate the uncertainties stemming from both the data and the models will be critical for the practical implementation of DL methodologies.
This challenging issue comprehends a topic currently under investigation in the DL community with no definitive answer. There are, however, multiple techniques that estimate uncertainty such as Monte Carlo dropout~\cite{gal2016dropout}, deep ensembles, and variational inference~\cite{abdar2021review}.
This uncertainty quantification will provide answers to the degree of confidence to which a model is working during normal operation. This will also entail the possibility to raise flags when a model is working out of its confidence interval, hence making it more robust and safer.

\section{Levels of automation}\label{sec:lev}

From the literature review, it can be seen that some works address single tasks (e.g. pre-processing or defect detection) while others encompass several (e.g. feature extraction and defect detection). However, the level of automation that they enable is not thoroughly discussed in these works (or any other), which can be detrimental for their practical implementation. This suggests that a clear automation path through different, well defined, levels can help categorise the DL contributions and establish the paths to increase the automation level in NDE.

\begin{figure}[h!]
    \centering
    \includegraphics{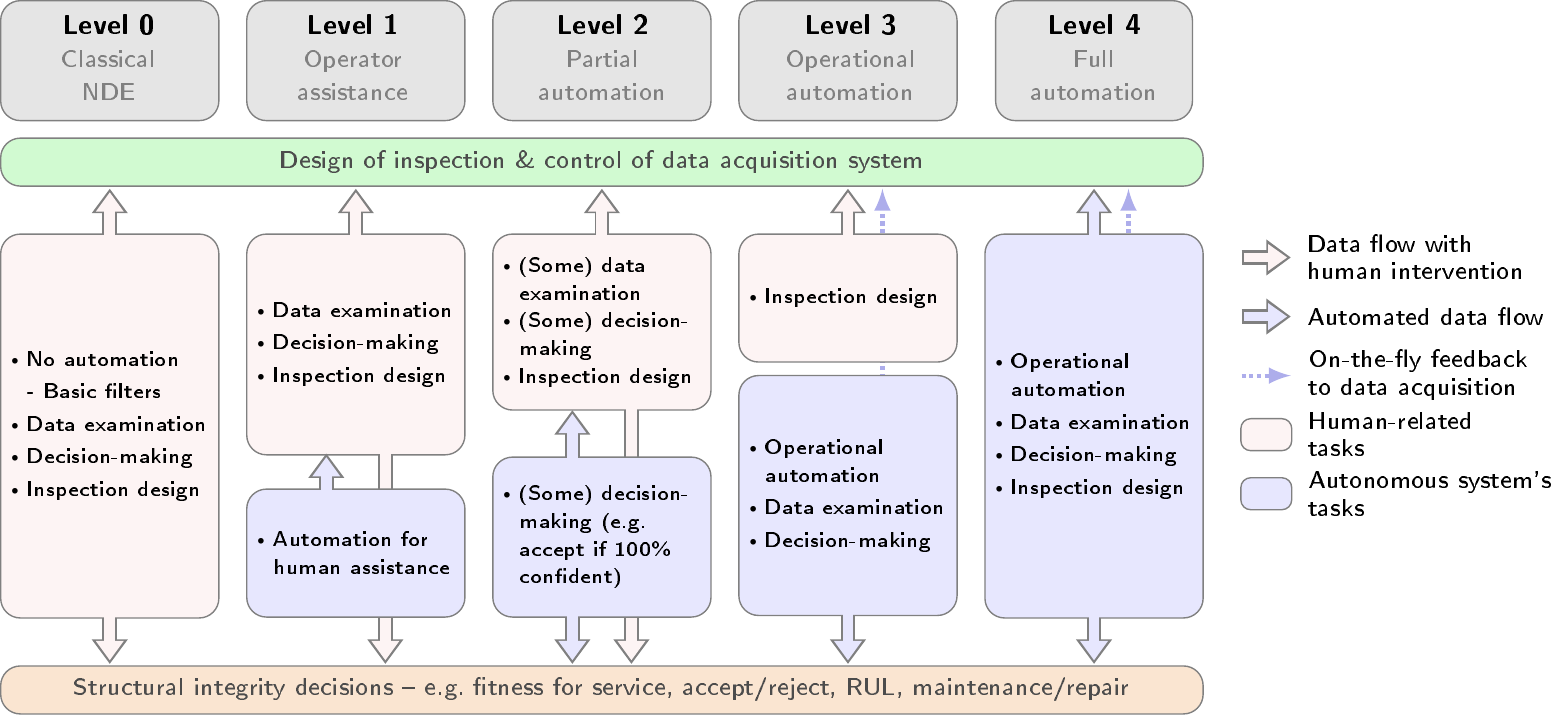}
    \caption{Diagram describing the proposed NDE automation levels and their associated responsibilities assumed by humans (in pink boxes) or autonomous systems (in blue boxes).}
    \label{fig:levelsDiagram}
\end{figure}

This section proposes and explores a series of levels of automation for any NDE modality, although the focus is primarily placed on ultrasound inspection. 
The levels are depicted in Figure~\ref{fig:levelsDiagram} and range from the classical NDE procedures, where the human operator is in charge of everything, to the ultimate future of full NDE automation with no human intervention. 
At Levels 3 and 4, operational  NDE is completely automated and the output can feed directly into structural integrity (SI) decision making. This in turn enables novel data-driven methods for SI decision making (e.g. accept/reject, maintenance/repair decisions, remaining useful life (RUL) estimation), which can also become completely automated.
Note that the proposed levels are in accordance with the recently published automation levels by the European Union Aviation Safety Agency (EASA)~\cite{EASA2020Roadmap,EASA2021concept}, i.e. level 1 of assistance to human, level 2 of human-machine collaboration, and level 3 of more machine autonomy. In the current paper, the levels are extended to take into account a wider range of applications around ultrasonic NDE.

One of the most important enabling factors of the transition from low to high automation levels is the development and implementation of DL-based approaches. These methods allow a higher degree of automation (i.e.~dealing with increasingly complex scenarios in a highly-efficient and accurate manner) that potentially requires less operator input and eventually should enable decisions to be taken without human intervention. Moreover, these methods are enabled by Industry 4.0 through the development of big datasets that, along with modelling data, will be used to boost the NDE automation with high confidence. Note that the necessity of big datasets poses a challenge due to the lack of experimental data nowadays as evidenced in the literature review. This issue, although temporary, can be tackled through the use of modelling data that complements the available experimental data, as discussed later and in some of the literature review. Although the availability of real unlabelled data is temporary, the issue of big, labelled datasets would remain a problem. This could be tackled using rudimentary DL systems to coarsely label data -- e.g. using object detection to coarsely label data for future segmentation model training~\cite{yang2010layered}. The following subsections explain the scope of the proposed levels and identify future opportunities and challenges that must be tackled in the context of DL-based NDE to achieve the potential of NDE automation.

\subsection{Level 0: Classical NDE}\label{sec:lv0}

The basic level of automation covers the classical NDE procedures, whereby all maintenance steps are manually addressed by a trained operator with basic support from data acquisition equipment but without any autonomous system. Only basic automation through filters defined by humans are considered at this level.

This level dates back to the early attempts at performing ultrasonic NDE.
Initially, multiple researchers developed a range of techniques and methodologies to detect defects and flaws within solid materials.
One of the first was by Firestone~\cite{firestone1946supersonic} in 1946 through the supersonic reflectoscope, a device able to carry out ultrasonic measurements in solid parts. This was a landmark that arguably established the beginning of ultrasonic NDE.
Additional contributions were made towards the improvement of quality assurance and detectability of defects, for instance, by detecting defects and measuring weld penetration lengths using wave propagation velocities obtained from pulse-echo tests~\cite{steffens1968nondestructive}.

It is worth highlighting that ultrasonic weld inspection was one of the most important topics for safety-critical industries and, as a result, a series of codes and recommendations were released~\cite{carson1972procedures} in the 1970s.
The main idea of these codes and standards for SI was to pre-perform SI calculations for generic structures so the inspection and sentencing requirements were already defined by the code.
Nonetheless, this was an open topic with no definite answer as to how the defect detection and characterisation should be performed to maximise reliability and accuracy of the inspections~\cite{deSterke1976advancements, young1979we}. To this end, several authors developed multiple contributions to better understand ultrasound-defect interactions and proposed mechanically-automated procedures to reduce the number of human errors.
For instance, the automation of an ultrasonic array probe to focus in different directions without operator intervention, with the directions having been pre-specified, was proposed in~\cite{iversen1976improved}; and the mechanisation of ultrasonic tests in pipeline weld inspections was also developed for more consistent results in~\cite{desterke1980automatic}. As a result of this research effort, defects could be detected, but accurate defect characterisation was still not feasible, leading to conservative sentencing~\cite{tomlinson1980ultrasonic}.
In general, ultrasonic NDE performance was limited by the operator's ability and hence prone to error. Several failures during operation due to undetected flaws that caused some catastrophic consequences (e.g. train derailments~\cite{garcia2006application} or engine failures due to undetected micro-fractures~\cite{remillieux2014review}) along with the surge in power and easy access to computational resources motivated a step forward in safety through the development of models and pre-processing algorithms. This enabled a substantial enhancement of the performance and a higher degree of automation, safety and reliability from ultrasonic NDE inspections.

\subsection{Level 1: Operator assistance}\label{sec:lv1}
Level 1 covers most of the existing literature and practise on ultrasonic NDE, especially on techniques and methods currently used by most industries for structural inspection. At this level, the operator is in charge of examining all NDE data, making all decisions and designing the inspections but with the assistance of computational/mechanical tools. If any autonomous system is used, it is to assist the operator (e.g. through image processing algorithms or computational inverse problems) rather than to replace the operator. 

Multiple techniques that allowed more complex and informative experiments were proposed with the initial use of ultrasonic arrays by adopting different configurations such as 1D, 2D, or circular arrays~\cite{drinkwater2006ultrasonic}.
Arrays enabled the use of focusing techniques such as the classical beamforming to increase the transmitted energy and hence reach larger inspection areas~\cite{olson2007beam, giurgiutiu2004embedded, cantero2020empirical}. Alternatively, the arrays could be used to fire elements independently while the rest of array elements are acquiring information, thus enabling the FMC~\cite{holmes2005post}. This can be arguably considered as the only experimental technique that can extract all the ultrasonic information from a probe array.
Note that all these techniques deal with big volumes of data and intrinsically automate their processing to convey, through image formation, interpretable data to the operator.

Data pre-processing is the ultrasonic NDE task that has arguably experienced the highest automation due, in part, to the fast computational development and the ease with which high-fidelity digital data can be acquired. In this context, techniques for the time- and frequency-domain analysis of ultrasonic data showed a larger evolution with the appearance of fast Fourier transform techniques, wave decomposition methods, or filtering approaches to reduce experimental noise~\cite{oruklu2009applications, pardo2006noise, song2006wavelet, lazaro2002influence, ruiz2005new}. 
FMC opened up the possibility of higher performance imaging algorithms, including ones that could adapt to surface profiles, deal with anisotropy, and generate multiple views using different wave modes (i.e. longitudinal or transverse)~\cite{holmes2005post, hunter2008wavenumber, zhang2014efficient, zhang2017investigation, zhang2010defect}.
More recently, the appearance of highly-capable CPUs, GPUs, TPUs (e.g. Google Coral), and edge-computing devices (e.g. NVIDIA Jetson) along with open-source software packages (e.g. TensorFlow~\cite{tensorflow2015-whitepaper}, Tensorflow Lite, and PyTorch~\cite{NEURIPS2019_9015}) has facilitated the development and deployment of DL methods for NDE as evidenced in the review of Section~\ref{sec:rev_proc}.

In the case of defect detection, several techniques have been typically applied, e.g.~thresholding techniques in time and amplitude domains~\cite{oruklu2009applications, song2006wavelet, zhang2010defect,cornforth1980ultrasonic}; probabilistic approaches that consider and quantify certain sources of uncertainty, such as aleatory or epistemic, have been used for anomaly detection applications~\cite{wu2017bayesian, fuentes2020autonomous}; and even comparing dispersion characteristics of guided-waves~\cite{masserey2014high}.
The characterisation of the defects has usually been addressed by visual interpretation of images assisted by physics-based models~\cite{bai2018ultrasonic, velichko2017ultrasonic, zhang2010use, zhang2008defect}. 
Inverse problems have been used to extract defect-related information (e.g. residual strength of damaged material) from the raw data based on physics- or data-based models~\cite{chiachio2017multilevel,bai2021use,sandhu2018bayesian}. These methods offer accurate and robust results, at the cost of more computational time than direct image interpretation. This drawback could certainly be addressed and eased with the use of DL within the inversion frameworks.

Despite being the most widespread automation level in practice, there is still scope to extend Level 1 automation to maximise the utility of DL-based operator-assisting techniques. These methods aim at enhancing the sentencing effectiveness of the NDE operator by providing more useful, easier-to-interpret NDE data. 
Three future opportunities or challenges are identified:
\begin{itemize}
    \item \textit{Cleaning raw NDE data}. This can encompass the suppression of random and coherent noise, including both micro-structural noise and signals from structural features such as such as reflections from the edges of the inspected part. It could enhance the defect detection rates given that false alarms due to artefacts present in the data would be minimised. In addition, the removal of non-defect data features could lead to more usable images with a larger usable area for both defect detection and characterisation. 
    \item \textit{Improving image interpretation}. There is an opportunity for DL to increase image resolution so that more accurate direct visual interpretation from ultrasonic images is possible. 
    Apart from the current contributions on image super-resolution, mainly focused on metallic parts, more types of defects and complex materials such as composite laminates could benefit from such a technique to provide accurate information about shape and extent of defects.
    \item \textit{Materials}. There are complex materials and structures, which are increasingly prevalent in some industries, and which pose unique data interpretation challenges -- e.g. composite materials. DL methods could be applied to address such data-related challenges.
    \item \textit{Operator-machine collaboration}. Human-machine collaboration could also be investigated with the operator rewarding good performance and penalising errors from the autonomous system. Provided that the interactions are well-controlled and that the autonomous system is sufficiently transparent for the operator, reinforcement learning techniques~\cite{mnih2013playing} (i.e. training method based on rewarding desired outputs and punishing undesired ones) could also enhance the capabilities of such a system, making it more effective even after its deployment.
\end{itemize}

\subsection{Level 2: Partial automation}\label{sec:lv2}

In Level 2, the operator addresses some of the data examination, its associated decision-making, and the complete design of the inspections. At this level, some of the decisions are made autonomously, e.g. accepting inspected parts if the system is highly confident. The end point of this level is further enhanced sentencing capabilities for NDE inspections enabled by the augmented performance of autonomous systems that partially substitute the NDE operator in basic scenarios designed by the operator.

The partial automation of NDE is another trending research topic in which powerful DL tools are exploited to detect and/or characterise defects from NDE data. However, their implementation in industrial applications is still limited to initial feasibility studies to understand the potential benefits and remaining challenges of this technology. For instance, Fujitsu developed and tested a DL approach for defect detection using data fusion and CNNs to improve the effectiveness and efficiency of quality inspections for wind turbine blades~\cite{furuya2019imagification}. Note that this type of automation approach needs to be scaled up and tested in fully-operational environments to ensure its accuracy and reliability in service. Nevertheless, a large number of contributions have been made from the research community demonstrating the potential of these techniques for achieving partially-automated NDE through defect detection and characterisation (see Sections~\ref{sec:rev_det} and~\ref{sec:rev_cha}). 

Relatively few DL methodologies have been proposed to combine multiple NDE steps, e.g., from data pre-processing to defect characterisation. 
The majority of the contributions that have been made are in the context of structural health monitoring and ultrasonic guided-waves. 
For instance, defect detection and characterisation were jointly addressed in a DL framework using a single convolutional network that gives all the defect-related output for ultrasonic guided-waves in aluminium plates~\cite{ewald2019deepshm}. Alternatively, the use of concatenated DL models have been explored for defect detection and characterisation using physical knowledge and ultrasonic guided-waves~\cite{rautela2021combined}. In general, having one or two separate networks for the automation of multiple steps has advantages and limitations. The training process and application of single DL models are more straightforward but at the cost of typically more complex model architectures, and hence large training databases. Alternatively, two independent, simpler, models provide a more controllable and interpretable/explainable scenario. The user can control and audit independently what the model is doing at any stage, which also makes this approach potentially better suited for regulated inspection qualifications. The intermediate output of this two-stage approach can allow a human to understand the decision-making process more easily~\cite{de2018clinically}.
Although both strategies can provide satisfactory results, it is recognised that the second approach (using different, concatenated, models) might have additional advantages from an industrial perspective: faster inspection qualification by having more control over the system or enhanced traceability in the interface between detection and characterisation models.

Despite the great research effort, DL-based partially-automated systems need improvement in terms of their generality, explainability, and overall adaptability to real-world structures outside laboratory environments. Therefore, the following points have been identified as potential future lines of work to enable and obtain the maximum performance from Level 2 of automation:
\begin{itemize}
    \item \textit{Explainability}. Detection and classification methods within partially-automated systems need to provide clear information as to how the results have been obtained. This is indeed not an easy task due to the black-box nature of DL models. However, coupling physics-based models and heuristic approaches that have been classically applied with newer tools will produce a greater transparency and control over the results. For example, this has been demonstrated by including the wave equation as an element to be satisfied in the loss function of a physics-informed neural network~\cite{raissi2019physics}. Partitioning the models to address smaller tasks that are more controllable and explainable is another candidate approach. \\Another important aspect of explainability comes from the fact that the autonomous system is partially substituting the operator who needs to supervise and audit the model outputs. The models need to give transparent information at all times of the outputs along with labels that are easily understandable by the operator. To this end, clear application programming interfaces (APIs) are essential elements of this automation level. 
    \item \textit{Generality}. The autonomous system needs to work under many varying conditions, which will require it to be robust and applicable to similar structures under different conditions. This can be achieved by using accurate physics-based models to generate training data, long-term data series, or even domain adaptation techniques (e.g. mixing model and experimental data for training). However, there is an evident limitation for some applications where only one type of data can be used for training (e.g. defect-free data). In these cases, unsupervised learning methods with thousands of measurements could be a viable option for autonomous defect detection as long as the model performance is demonstrated with data from multiple lab and industrial trials. 
    \item \textit{Uncertainty quantification}. In line with the previous points of explainability and generality, new models should be accompanied by a thorough analysis of their associated parameter uncertainties in several circumstances, especially in the case of defect characterisation DL models. The distinction between aleatoric and epistemic uncertainties will also provide an idea as to how accurate and general a DL model is and what are the associated difficulties.
    \item \textit{Uncertainty propagation}. Level 2 potentially entails the connection of independently designed models for the systematic functioning of the autonomous system. This has some challenges associated with it that need to be addressed, e.g. how the models are going to inherit information from lower levels and how uncertainties are going to propagate through.
    This is an incipient area and one from which industries using complex structures (e.g. from aerospace or wind energy sectors) could benefit.
    \item \textit{Traceability and auditability}. It is important that any automation system enabled by DL is traceable and auditable. Even in the event that the output of the system is not fully explainable, the system should provide information to trace the model architectures, the implementation version, or the training data that was used for developing the DL model. In this context, joint collaboration between researchers and NDE industries is key to achieve a successful traceable system, which can be audited at some point in the future. 
\end{itemize}

\subsection{Level 3: Operational automation}\label{sec:lv3}

The end objective of Level 3 is to achieve the full automation of the NDE process. To this end, the autonomous system examines all data and makes all decisions for a given inspection scenario, thus enabling a complete operational automation. The NDE process is sufficiently automated that it can be used to provide on-the-fly feedback for optimal placement of data acquisition systems. The human operator is needed to design the inspections carried out autonomously and supervise the system. The autonomy that Level 3 enables can create a direct link to SI assessments without human intervention, e.g. deciding fitness for service or the optimum moment for maintenance or repair actions based on RUL predictions. 

It is worth noting that very limited contributions have been made in the context of data-driven operational automation mainly due to the lack of data, technology, methodological developments, and updated regulatory frameworks.
Another factor in this scarcity is that some of the elements needed for operational autonomous systems are not really amenable to scientific research (e.g. development of APIs).
Nevertheless, a few examples that have the potential to allow complete operational automation can be found in the literature for techniques other than ultrasound. For instance, Park \emph{et al.}~\cite{park2020concrete} have illustrated the automation of a laser-based system for the detection and characterisation of cracks in concrete structures. Unmanned aerial vehicles (UAVs) are used as a source of automated data acquisition, whose data are used for the detection of cracks with CNN-based DL models, while the characterisation of cracks are addressed using linear regression. Additionally, Xu \emph{et al.}~\cite{xu2019wind} have addressed the inspection of wind turbine blades through images taken by an UAV. In this case, a CNN-based model is used for both detection and characterisation simultaneously by creating different defect classes ranging from no defect to different types of defect (e.g. surface corrosion).
However, no DL-based methodologies or examples for this automation level have been provided for ultrasonic NDE inspection.

The analysed works show the potential of DL techniques in automating the NDE process. However, a more integrated approach that includes all the elements in the context of ultrasonic NDE inspection using autonomous systems is not yet available either in industry or the research communities. In addition to the challenges identified for the previous levels, the following are proposed as future work needed to address in order to achieve maximum performance and reliability from an autonomous system in Level 3:
\begin{itemize}
\item \textit{Outlier identification}. The automation of the inspection system that is designed by an operator has to account for scenarios that are outside the model's confidence intervals and be able to raise a flag. In this context, outlier identification techniques should be embedded within the autonomous system so the operator can take over the situation and override the autonomous system in order to keep high safety standards.
\item \textit{Modularity, integration, and communication}. Level 3 may include robotic systems that are controlled by an autonomous system. This poses complex modularity, integration and communication issues stemming from the fact that information needs to be passed both ways -- robotic to autonomous system and system to robot. For example, in robotic acquisition systems, the scanning paths can be based on the analysis of the data acquired.\\ Another important element is the communication and interoperability of the autonomous system with its subelements (e.g. robots and submodels). This communication needs to be robust and unambiguous, and should also be traceable at all times. Therefore, the development of APIs would need to extend and integrate the robot-autonomous system communications and the rationale by which the decisions are taken (e.g. stop taking measurements).
\item \textit{Optimal sensor/robot placement}. The autonomous system is responsible of taking measurements in the inspected structure. In some scenarios this will include the decision on where to locate the robot to obtain the best data that minimises the uncertainty of the related defect detection and characterisation. In these cases, optimal sensor or robot placement approaches will have to be developed so that the autonomous system can correct the robot in real time and move it towards the optimal location. The development and integration of these optimisation techniques within DL frameworks is identified as a key element for successful operational automation.
\end{itemize}

\subsection{Level 4: Full automation}\label{sec:lv4}
At this level, everything in the inspection-maintenance process is done by the autonomous system (inspection design, measurements, processing, NDE, and decision-making) in all scenarios. The NDE operator or SI practitioner would not be needed anymore. Assistance to the autonomous system would be required if deemed necessary for maintenance procedures of the system or to proceed with maintenance or manufacturing decisions taken by the system. 

Level 4 remains aspirational at this point in time, but the automation tools required are beginning to be reported~\cite{syed2020robotic, goel2020robotics} and the path already initiated points to the full automation of repetitive and predictable processes. It is worth noting that despite the similarities with Level 3, the required certifications and qualifications of the autonomous system would be much stricter, with almost perfect levels of accuracy in all possible inspection scenarios with all the potential damage, materials, and environmental and experimental conditions that can appear. Technology should provide, at this stage, even more accurate DL techniques which can be easily implemented in different devices. All the model outputs need to be extremely clear and saved carefully with appropriate labels so that future audits can be performed on the autonomous systems by trained operators or regulators. Maintenance- and manufacturing-related decisions are taken within the system with physics-informed data and the operator or other intelligent system should be able to address these tasks.

There are, however, some remaining challenges that have not been mentioned in the previous levels that concern the full automation of NDE processes:
\begin{itemize}
    \item \textit{Retaining know-how}. A potential issue of fully-autonomous systems is that the industries may lose human-related know-how, leaving the autonomous systems with all the knowledge for NDE and structural health management. Despite no human operators being needed in the loop, the physics-based knowledge and the functioning principles of the autonomous system needs to be known by inspectors and regulators that interact with the autonomous system -- e.g. to handle an alarm due to uncertainty. Therefore, new, standardised, procedures for stating the underlying physics and the DL models functioning will need to be developed.
    \item \textit{Reliability}. The quantification of the autonomous system reliability is a critical element of the full automation. The systems need to be completely autonomous and hence their confidence levels sufficiently high to be independent of human intervention. Methodologies for quantifying reliability while considering diverse uncertainties are therefore key for this stage.
\end{itemize}

\subsection{Mapping of contributions into the proposed levels}
\begin{figure}[htb]
    \centering
    \includegraphics{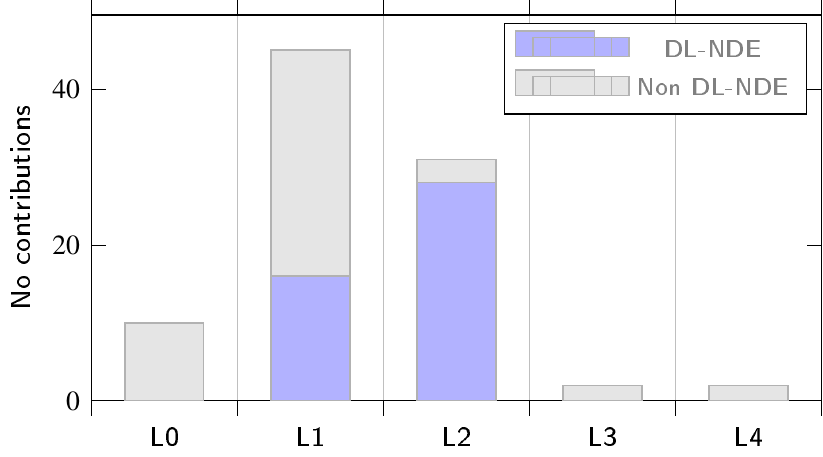}
    \caption{Scientific contributions accounted in this review mapped into the proposed automation levels.}
    \label{fig:map}
\end{figure}

An approach to identifying the overall state of research is to map the DL and non-DL contributions onto the proposed levels of automation. As shown in Figure~\ref{fig:map}, all DL-related contributions are at Levels 1 and 2. Note that the papers defined as Level 1 are those dealing with data pre-processing (Section~\ref{sec:rev_proc}), while the contributions in Level 2 are those stemming from defect detection (Section~\ref{sec:rev_det}) and characterisation (Section~\ref{sec:rev_cha}) as well as property measurement (Section~\ref{sec:rev_prop}). Note that this review is not exhaustive but provides a global picture of the most important contributions in DL-based ultrasonic NDE. The rest of the grey columns in Figure~\ref{fig:map} accounts for the papers cited along the description of the levels of automation in the above sections; hence not directly related to DL or NDE. This analysis reveals that most of the effort on DL-based NDE is being put into enabling a partial automation through defect detection and characterisation as well as property measurement. This is possibly as a result of the large number of available DL models (exemplified in Section~\ref{sec:dl}) that are readily available to be implemented in multiple applications, especially dealing with images as input data. Nonetheless, there is still a long way to go until these two levels are fully addressed and exploited by the NDE community. Higher automation levels enabled by DL are aspirational at this point in time, with the methodology development in Levels 1 and 2 still in its infancy.

\section{Basic axioms for DL-based ultrasonic NDE}\label{sec:axi}
Having reviewed the state-of-the-art and the contributions made at each automation level, it is easy to appreciate how heterogeneous the current DL methodologies are from paper to paper. For instance, some authors propose models that address different steps simultaneously~\cite{ewald2019deepshm} while others do it with independent models~\cite{rautela2021combined}. Explicit model limitations (e.g. range and sensitivity of hyper- and pre-processing parameters) are typically left out of the papers, hence giving partial information about the use and practicality of the DL models. These inconsistencies and lack of information cause slower advances in the automation path as industries and regulators may need to see a more unified approach. A series of \emph{axioms} are proposed below to standardise the process of developing and applying a new DL model for NDE and bridge the current gap between the multiple DL contributions and the industrial realm, where regulations and inspection qualifications are essential. It is worth highlighting that the axioms have been discussed with multiple industrial contacts across a range of sectors including aerospace, nuclear and renewable energies and are in line with the current expectations of reports from the EASA~\cite{EASA2020Roadmap, EASA2021concept} and the European methodology for qualification of NDE~\cite{eniq31}. Moreover, these axioms (summarised in Table~\ref{tab:axioms}) build on the technical evidence found in the literature and the future challenges identified within each automation level.

\begin{axiom}
Deep learning models need to be trained on a predefined domain of operation.
\label{ax:dom}
\end{axiom}
\noindent DL models are expected to demonstrate high performance within a predefined domain of operation, from which training and validation data are obtained. It is currently unrealistic to try to design a model that works for all conditions, structures and techniques. Therefore, the DL models for NDE must have a clear scope depending on the NDE modality, the types of structures and materials under inspection, and environmental changes expected. This axiom is also important for inspection qualifications in order to assess the performance of a given DL model in the appropriate context.

\begin{figure}[h!]
    \centering
    \subfloat[]{
    \includegraphics{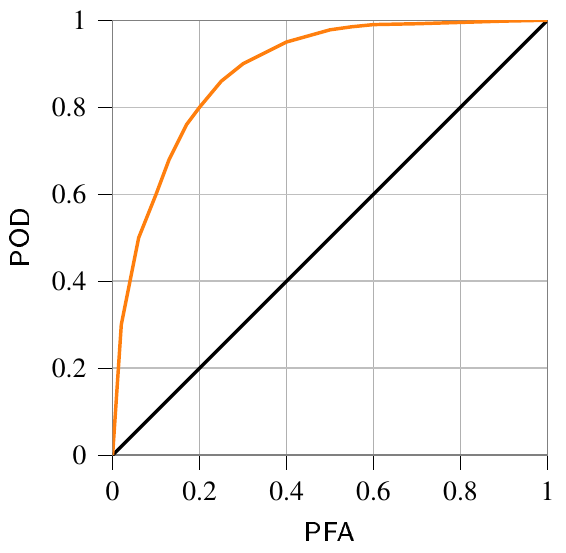}\label{fig:ax1_exA}}
    \subfloat[]{
    \includegraphics{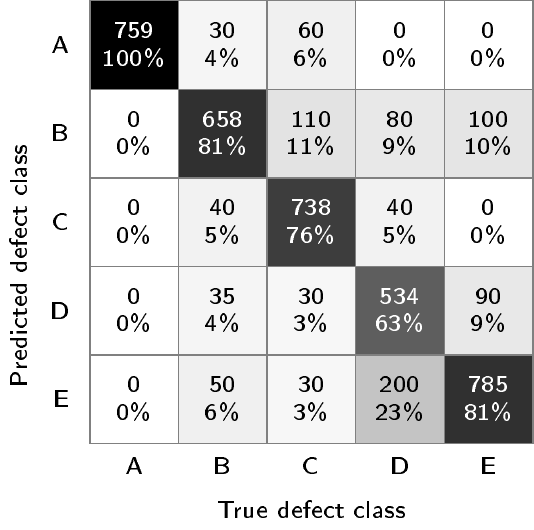}\label{fig:ax1_exB}}
    \caption{Examples of (a) ROC curve and (b) confusion matrix for the evaluation of detection and classification methods, respectively.}
    \label{fig:ax1_ex}
\end{figure}

\begin{axiom}
Deep learning models need to have quantifiable performance across the predefined domain of operation.
\label{ax:perf}
\end{axiom}
\noindent DL models need to be sufficiently \emph{performant} depending on their application (e.g. detection, characterisation, pre-processing) so that the required levels of effectiveness informed by the norms and regulations are satisfactorily met. To this end, their performance in addressing these tasks needs to be quantified across the entire domain of operation by appropriate means depending on the end objective of the DL model. For instance, models designed for detecting defects may have probability of detection, probability of false alarm, or receiver operating characteristic (ROC) curves~\cite{nockemann1991reliability} (see Figure~\ref{fig:ax1_exA} as an example). If the characterisation of defects is the end point of the DL model, then continuous indices measuring the discrepancy between predicted and real properties may be adopted. In cases where the type of defect needs to be provided, classification indexes can be adopted such as confusion matrices (see Figure~\ref{fig:ax1_exB} as an example) and reliability diagrams for Bayesian neural networks~\cite{kuleshov2018accurate}.

\begin{axiom}
The uncertainties related to the deep learning model, data, and hypotheses need to be quantified.
\label{ax:uq}
\end{axiom}
\noindent The quantification of uncertainty is a critical element for DL models that are built to work autonomously. This will provide relevant information about the confidence levels of the models in their working environment. To this end, Bayesian techniques for the quantification of uncertainty can be used directly within the construction of DL models, e.g. through Monte Carlo dropout, Markov chain Monte Carlo, variational inference and Bayesian active learning~\cite{abdar2021review}. If the models are designed for automation levels equal or higher than Level~2, the effect of the propagation of uncertainties through multiple chained models also needs to be quantified. This will be an essential element to quantify and possibly mitigate the effects of uncertainty amplification if errors are propagated from the initial stages of the NDE process. For Level~3 and up, this information will be required to let operators know when their input is required.

\begin{axiom}
The values of any adjustable deep learning model parameters must be specified.
\label{ax:pars}
\end{axiom}
\noindent Another critical element for the inspection qualification is the identification of influential and essential parameters of an inspection method~\cite{eniq31}. When automation is enabled by autonomous systems, these parameters are likely to be related to model pre-processing parameters that affect the performance of the trained models such as scaling or delaying parameters applied to the input data. To this end, parametric studies can be applied for the quantification of the output variations depending on the input variation, such as global sensitivity analysis~\cite{saltelli2008global}. The model hyper-parameters can also be adjustable in post-training use cases of model compression techniques (e.g. knowledge distillation~\cite{gou2021knowledge}, quantisation~\cite{carreira2017model}, and model pruning~\cite{zhu2017prune}) prior to model deployment -- i.e. to develop efficient DL models from larger ones and enable them on edge devices~\cite{chen2019deep}.

\begin{axiom}
Deep learning based autonomous systems need to be self-aware of their own limitations.
\label{ax:awa}
\end{axiom}
\noindent Autonomous systems in Levels 3 to 4 must be able to independently identify input data that is out of the designated domain of operation (see Axiom~\ref{ax:dom}). This can be addressed by quantifying the epistemic uncertainty of the DL model output, e.g. if the model is uncertain about its prediction, then the input data are likely to fall out of the domain of operation. There are two immediate consequences of this axiom: (1)~the system is more independent of the operator by needing less input in order to assess new data, and (2)~the data that is flagged by the autonomous system can be used for later analysis and enhancement of the DL models. Risk assessment of these scenarios and mitigation plans that foresee how the flagged data are going to be addressed (such as having the operator taking over the NDE process) should also be in place for the autonomous system to be approved by relevant regulatory bodies.

\begin{axiom}
Deep learning models within an automated autonomous system need to be traceable.
\label{ax:trac}
\end{axiom}
\noindent At high automation levels, especially when the autonomous system has the main responsibility in providing answers to each NDE step (and also to SI), the traceability of the results in the interface of models is essential. APIs could be designed to produce the appropriate log files with informative labels and pointers to the data location. This entails the possibility to trace back output data with explanatory labels and metadata from the latest stages of health management and maintenance decisions, to the initial stages of data acquisition. 
Traceability will also be important for (1)~industries to keep the know-how of autonomous systems through informative metadata; and (2)~regulators that conduct audits to ensure that the autonomous decisions are taken appropriately.
Another potential scenario where the autonomous system could benefit from traceable information is to carry out corrections and/or internal optimisation of the NDE process. For instance, the system may be able to check that the final output is out of its confidence intervals and go back to try to optimise the location for acquiring NDE data so that the output is less uncertain.

\begin{table}[width=1\linewidth,cols=3,pos=h]
\caption{Summary of the DL-based NDE axioms along with their elements and proposed approaches to address them.}\label{tab:axioms}
\begin{tabular*}{\tblwidth\footnotesize}{@{} LLL@{} }
\toprule
No& Description & Proposed techniques \\
\midrule
Axiom 1 & \textit{Deep learning models need to be trained on a predefined} & Assessing scope of application\\
 & \textit{domain of operation} & \\
\midrule
Axiom 2 & \textit{Deep learning models need to have quantifiable} & ROC curves~\cite{nockemann1991reliability} \\
 & \textit{performance across the predefined domain of operation} & Reliability diagrams~\cite{kuleshov2018accurate}  \\
\midrule
Axiom 3 & \textit{The uncertainties related to the deep learning model,} & Monte Carlo dropout~\cite{gal2016dropout} \\
 & \textit{data, and hypotheses need to be quantified} & Variational inference~\cite{swiatkowski2020k} \\
 & & Bayesian active learning~\cite{gal2017deep} \\
\midrule
Axiom 4 & \textit{The values of any adjustable deep learning model} & Global sensitivity analysis~\cite{saltelli2008global}\\
& \textit{parameters must be specified} & Model compression~\cite{chen2019deep} \\
\midrule
Axiom 5 & \textit{Deep learning based autonomous systems need to be} & Epistemic uncertainty~\cite{huang2021quantifying}\\
& \textit{ self-aware of their own limitations} &\\
\midrule
Axiom 6 & \textit{Deep learning models within an automated autonomous system} & Logging DL model results\\
& \textit{need to be traceable} & Saving input data \\
\bottomrule
\end{tabular*}
\end{table}

\subsection{Illustrative examples}
The proposed axioms (summarised in Table~\ref{tab:axioms}) address the main technical areas of interest in order to make any newly developed DL-based inspection procedure realistic and industrially applicable in the near future. 
The authors expect that the proposed axioms become a pseudo-checklist against which the NDE practitioners and research community can quickly monitor the viability of a newly-developed DL model. 
In this context, it is understandable that not all axioms have been addressed in the previously reviewed papers as most of them entail proof of concepts and small, yet important, advances to the technology. Nevertheless, it is desirable to explore the level of adherence of the latest papers to the proposed axioms.

\begin{figure}[h!]
    \centering
    \includegraphics[width=.8\textwidth]{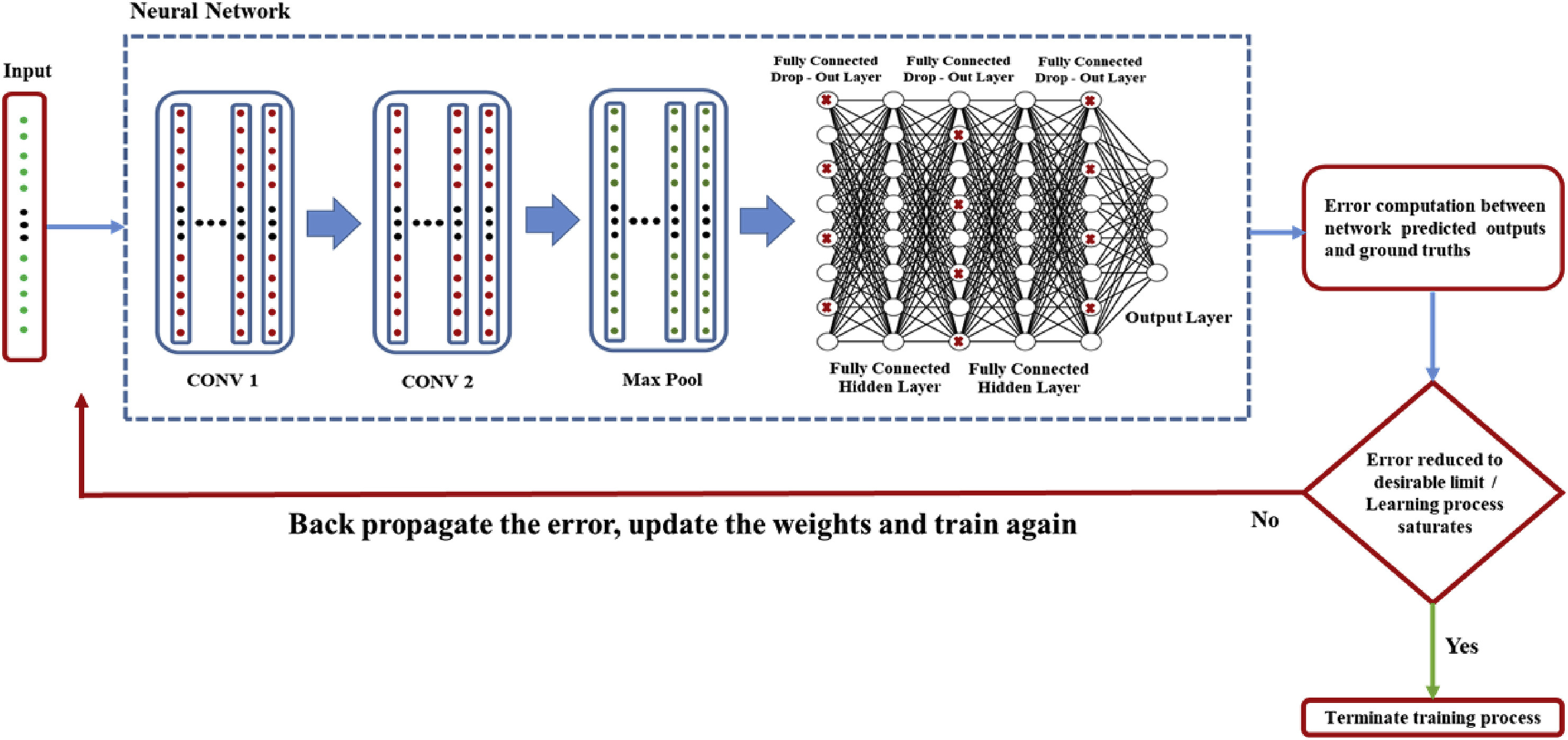}
    \caption{Classification network architecture for weld defects developed by~Munir \emph{et al.}~\cite{munir2020performance}}
    \label{fig:munir}
\end{figure}

Let us analyse the evolution of the publications made by the Department of Mechanical Engineering in Sungkyunkwan University (South Korea) that have been authored by Munir \emph{et al.}~\cite{munir2018investigation,munir2019convolutional,munir2020performance} as a case study. In all of them, the authors address the classification of defects in welds using DL methodologies. In the 2018 publication~\cite{munir2018investigation}, a FCNN was proposed and Axiom 2 about quantifiable performance was addressed by giving the accuracy level of the trained model. No indication about the domain of operation (Axiom 1) was provided, i.e. no information regarding the range of crack lengths and angles or porosity density, and this significantly limits the applicability of such an approach. Later, in the following works~\cite{munir2019convolutional, munir2020performance}, the authors proposed the use of a CNN model for the classification of defects, whose architecture is depicted in Figure~\ref{fig:munir}. In these papers, the authors addressed clearly the performance of the models (Axiom 2) through the quantification of accuracy levels, gave indication about the dispersion (or expected variance given the selected hyper-parameters) of the accuracy based on the repetition of the training process (Axiom 3), and gave a brief description of the domain of operation (Axiom 1). This series of works show a progression in line with the first three proposed axioms.

\begin{figure}[h!]
    \centering
    \includegraphics[width=.6\textwidth]{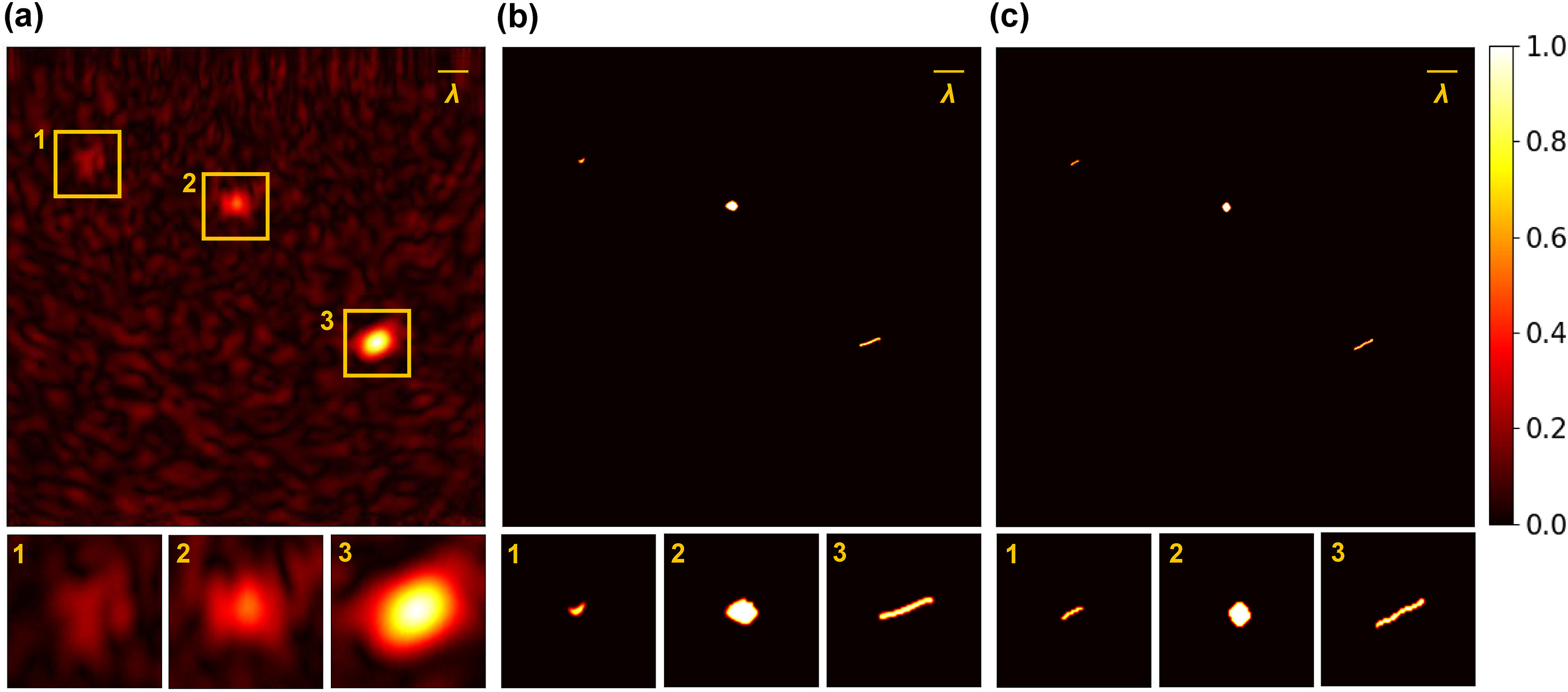}
    \caption{Qualitative performance quantification by image comparison from Song \emph{et al.}~\cite{song2020super}. Original view in (a), reconstructed image in (b), and ground truth in (c).}
    \label{fig:song}
\end{figure}

In the paper published by Song \emph{et al.}~\cite{song2020super}, the authors addressed the formation of high-resolution ultrasonic images to characterise defects accurately. They included a description of the domain of operation (Axiom 1), whereby the different types of defects contained in the training dataset were provided. The performance of the model (Axiom 2) of the model was not quantified and only a qualitative comparison with the ground truth images are given (see Figure~\ref{fig:song}), which significantly limits an objective analysis of the approach.
In a similar situation, the works by Virkkunen \emph{et al.}~\cite{virkkunen2021augmented} and Park \emph{et al.}~\cite{park2021porosity} address the axioms related to the quantification of performance (Axiom 2), e.g. by probability of detection (PoD) curves and confidence matrices, as well as a description of the domain of operation (Axiom 1). However, there is no information about the uncertainty of the models (Axiom 3). Another interesting example is the defect (cracks) characterisation (length and angle) CNN model proposed by Pyle \emph{et al.}~\cite{pyle2020deep} based on ultrasonic images. In this work, the performance of the model (Axiom 2) was quantified using the error between predicted and real data, the quantification of uncertainty (Axiom 3) was partially addressed by training the model 80 times using different initial points and obtaining the dispersion values, and the domain of operation (Axiom 1) was clearly defined by specifying the parameters used in the training database. 
In a recent work by Pyle \emph{et al.}~\cite{pyle2022uncertainty}, the authors addressed the notion of uncertainty quantification (Axiom 3) in DL-based NDE. The epistemic uncertainty of a CNN for sizing pipe cracks from ultrasonic images (initially proposed in~\cite{pyle2020deep}) was quantified through different methods, namely, deep ensembles and Monte Carlo dropout. The success of these methods was evaluated through their calibration and anomaly detection performance. Deep ensembles provide the best results although the quantified uncertainty tends to underestimate error.

Most of the reviewed works address the quantification of the performance and the description of the domain of operation, with some considerations of uncertainty quantification, i.e.~Axioms 1, 2, and 3. However, there is no clear consensus on the methods used to satisfy these axioms, especially axioms 2 and 3. For Axiom 2, many works rely on the direct estimation of model accuracy or error levels, but do not directly evaluate the DL model in an inspection environment, e.g. through PoD curves and ROCs. The quantification of uncertainty (Axiom 3) is also heterogeneously addressed with some authors training the model multiple times while others quantify probability density functions of the inferred parameters.

In this context, it can be observed that there are some important factors that are not addressed in most of the reviewed papers (due to the current, early, stage of this technology) that must be in order for them to be applicable to industrial settings. 
These are related to (1)~the description and identification of influential and essential parameters -- Axiom 4; (2)~the definition of standardised procedures to make the autonomous systems self-aware when they are working outside the domain of operation -- Axiom 5; and (3)~the traceability of multiple models assembled in autonomous systems -- Axiom 6.
These elements are essential for the real-world application of this technology and are critical for DL methods to adopted in industrial applications; but they still require a response from the scientific community. These three additional axioms have partially been identified thanks to the feedback provided by industries from multiple sectors (e.g. aerospace, oil\&gas, or nuclear). It is worth highlighting that Axiom 6 is more related to the implementation and development end, whereby the DL models are developed and an assembly software platform is needed. In this sense, addressing this axiom is more in the domain of software engineering, systems integration, and operational procedures than scientific research, but it should be considered at development stages in any case.

\section{Conclusions and future directions}\label{sec:conc}
This paper has reviewed the current state of research of DL-based NDE. 
DL has been successfully used in many applications, such as computer vision or natural language processing. For NDE, DL is principally being used as a way to: (1) perform data processing tasks that were either too slow or not possible with classical approaches; (2) create inspection procedures that are more independent of the operator's experience (hence less prone to errors) -- e.g. for damage characterisation; and (3) to automate repetitive NDE tasks such as defect detection from complex data and structures. However, these works show a great heterogeneity in the process of developing the DL models, possibly posing practical limitations on their applicability to real (industrial) settings.

A NDE automation roadmap has been proposed that presents a series of levels of automation ranging from fully operator-driven NDE to fully-automated NDE and SI. The levels, which are based on those from other industries such as the aviation sector, delimit the expected obligations of both human operator and autonomous system at different stages. It has been found that the current state of research is mostly concentrated around Levels 1 and 2, i.e. operator assistance and partial automation respectively. A series of open challenges associated with each level are also identified and described. Note that the aim of this roadmap is to set a horizon to both researchers and NDE industries so that the objectives for stepping up to a higher level of automation can be more easily set; hence standardising the automation journey.

In this context, the most immediate future opportunities associated with challenges that remain open for the research community have been identified:
\begin{itemize}
    \item \textit{Focused data denoising}. Most research is currently focused on denoising raw ultrasonic signals, but the focused identification of elements of interest (e.g. ToF of defect echoes or separation of overlapping echoes) remains only vaguely explored. This type of data manipulation could drastically improve the PoD while providing fewer false alarms. Another potential advantage of using cleaner data are easier and more accurate defect characterisation by NDE operators. DL could help through methods that automatically separate echoes, provide ToF information, even in the presence of heavy noise. 
    \item \textit{Image interpretation}. While there are many existing methods to create and interpret ultrasonic images available in the scientific literature, these can be limited for an accurate characterisation of small defects. DL super resolution algorithms are being explored with remarkably accurate results. The further development of these methodologies for materials with complex inner structures (e.g. composite materials) could certainly be a game changer by providing accurate images that are more easily interpretable for NDE operators.
    \item \textit{Uncertainty quantification}. The use of DL in real inspection scenarios is subject to the evaluation of the confidence intervals of the used models. To this end, a critical step is to quantify uncertainties in DL models that specifically measure both errors related to (1) the aleatory nature of experimental measurements and (2) the irreducible uncertainty associated with the models themselves. This information will be critical for the refinement of models until they are both accurate yet reliable (i.e. avoiding large output variation). Despite the criticality of uncertainty quantification, it remains scarcely addressed in the literature, and hence it is anticipated that it will become one of the most researched in the future in the field of DL-based NDE.
    \item \textit{Automated system self-awareness}. DL-based autonomous systems that are intended to automate tasks, must also be able to identify scenarios that are outliers to their pre-defined domain of operations. When an outlier is identified, the system should raise a flag either for the operator to take over or the designer to re-train or re-design the DL models. To this end, one possible approach is through the quantification of epistemic uncertainty, whereby the system can tell itself that it is not confident enough in the recent prediction. It is also anticipated that the development of self-awareness methodologies for DL-based NDE models will be an essential research topic in the foreseeable future.
\end{itemize}

Lastly, to homogenise the development of DL-based inspection methods, this paper also identifies for the first time the axioms that DL methods should satisfy to be fully applicable to NDE. The proposed fundamental properties are focused not only on the definition and evaluation of the DL models (e.g. quantifiable performance or uncertainty quantification), but also include implementation aspects such as the traceability. It is important to highlight that these axioms are derived from the literature review with input from diverse industries.

\section*{Acknowledgements}
The work reported is part of a pilot project (Grant number 100374) funded by Lloyd’s Register Foundation and the Alan Turing Institute Data-Centric Engineering Programme. The authors would also like to thank the industrial participants who responded to the surveys from which the axioms and automation levels were developed.

\printcredits

\bibliographystyle{model1-num-names}
\bibliography{DL_UNDT}

\end{document}